\documentclass[aps,amsmath,amssymb,showpacs,pre]{revtex4-1}
\usepackage{graphicx,color}
 \graphicspath{{./}{./figures/}}
\usepackage{ams math}
\usepackage{enumerate}
\usepackage{mathbbol}
\usepackage{amsfonts}
\usepackage{natbib}

\usepackage{bm}
\usepackage{hyperref}

\usepackage{color}

\begin{document}

\title{The role of non-conservative scattering forces and damping on  Brownian particles in optical traps}
\author{Matthieu Mangeat, Yacine Amarouchene, Yann Louyer, Thomas Gu\'erin and David S. Dean}
\affiliation{Univ. Bordeaux, CNRS, LOMA, UMR 5798, F-33400 Talence, France}

\bibliographystyle{apsrev}

\begin{abstract}
We consider a model of a particle trapped in a harmonic optical trap but with the addition of a non-conservative radiation induced force. This model is known to correctly describe experimentally observed trapped particle statistics for a wide range of physical parameters 
such as temperature and pressure. We theoretically analyse the effect of non-conservative force on the underlying steady state distribution as well as the power spectrum for the particle position. We compute perturbatively the probability distribution of the resulting non-equilibrium steady states for all dynamical regimes, underdamped through to overdamped and give expressions for the associated currents in phase space (position and velocity). We also give the spectral density of the trapped particle's position in all dynamical regimes and for any value of the non-conservative force. Signatures of the presence  non-conservative forces are shown to be particularly strong for in the underdamped regime at low frequencies.
\end{abstract}

\maketitle
\section{Introduction}
The seminal work of Ashkin in the 1970s instigated the development of optical traps \cite{ash}.
Optical traps or tweezers are an invaluable tool for studying statistical mechanics, both in an out of equilibrium, at the level of single particles. Moving the center of the optical trap allows one to measure very small forces in a wide range of systems, including colloids \cite{colloid}, biological systems (for instance viruses, proteins and  biopolymers) \cite{bio}, as well as  dielectric and metallic nano-particles \cite{nano}. The same kind of perturbation can be used to analyze non-equilibrium processes, for instance optical tweezers have been used to demonstrate various non-equilibrium fluctuation theorems  \cite{fluct}. However in such studies the optical trap does not simply furnish  a simple potential, optical scattering forces actually generate a non-conservative component to the overall force exerted by the trap. Because of the presence of these scattering forces, even static optical traps represent non-equilibrium systems as the particles trapped within them are driven by the non-conservative forces and thus are not described by a Gibbs-Boltzmann probability distribution. For instance, it has been demonstrated \cite{grier2008,grier2009,grier2010,grier2015} both theoretically and experimentally that scattering forces induce non-equilibrium currents, so called Brownian vortices, for trapped overdamped Brownian particles. At atmospheric pressures and near room temperatures, the underlying dynamics of a trapped particle can be taken to be overdamped Brownian. However optical traps operated at low pressure can probe dynamics where inertial effects become important and cross over  to the underdamped regime. Clearly the current state of the art of the theory has to be modified to take into account the effects of inertia and this is the goal of this paper. Using the model of a Gaussian optical trap in the presence of a simplified non-conservative force we examine the impact of inertial effects across the whole regime between overdamped and underdamped Brownian motion. 

First a general theory for the probability distribution of the perturbed steady state due to a small non-conservative force is developed for general damped Brownian dynamics. The modification of the steady state probability distribution function for the full phase space in the optical model of \cite{grier2015} is found and the resulting marginal probability distributions for the particle position and velocity is computed. We demonstrate that the Brownian vortices found in the current of the particle position persists and remains geometrically the same as that found in \cite{grier2015}. However the amplitude of the vortices depends on the damping parameters, we will show that this damping parameter can be tuned in order to optimize the non-equilibrium current. In addition, we show that there is a non-equilibrium current in the space of velocities, the vortex here turns out to be remarkably similar to the vortex in position space.

For the optical trap model of \cite{grier2015} we also compute the spectral density of the particle position for general damping. Our result recovers the computation of \cite{laporta2011} for overdamped Brownian motion and generalizes it to all damping regimes.  
Here we see that the non-conservative force generates very strong changes in the spectral density at low frequencies and that the amplitude of the low frequency spectral density can be tuned by varying the friction coefficient $\gamma$ and that there is a critical value of $\gamma$ where it is minimized. Finally, we also derive a function correlation which is a signal of the breaking of time reversal symmetry due to the non-conservative optical scattering force.

Some of the results derived in this paper are used in a accompanying letter, where
the role of inertia in the presence of non-conservative forces is demonstrated in the experimental context \cite{letter}. In this paper we concentrate on the presence of 
a nonconservative force in addition to a purely harmonic trap. This simplified model 
allows the derivation of a large number of analytical results. However, to fully explain the 
results of \cite{letter}, where small anharmonicities can give rise to large effects in the underdamped regime due to near resonance phenomena, quartic terms in the 
potential must be considered to fully reproduce the experimental results. However the 
results derived here can be adapted to the experimental context by using effective harmonic parameters describing the nonlinear system \cite{Renz1}. This therefore yields an analytic theory sufficient to completely describe the experiments and fully non-linear numerical simulations \cite{letter}.

\section{Modeling particles in optical traps}
\subsection{Langevin Dynamics}

In this section we first discuss the underlying Langevin dynamics used to model the trapped particle. We then discuss the underlying model for trapping,  aimed at experts in non-equilibrium statistical mechanics who are perhaps not familiar with the 
physics of optical trapping.

In this paper we will consider the dynamics of Brownian particles in an optical trap. If ${\bf x}(t)$ denotes the particle position, the underlying Langevin equation is
\begin{equation}
m\frac{d^2}{dt^2} {\bf x}(t) = -\gamma  \frac{d}{dt} {\bf x}(t) +{\bf F}({\bf x}) +\sqrt{2T\gamma}{\bm \xi}(t),\label{lan1}
\end{equation}
where $m$ is the particle mass, $\gamma$ the friction coefficient and ${\bf F}({\bf x})$ is a time independent force generated by the optical trap. The noise term is assumed to have Gaussian white noise correlations in time, this noise has zero mean and correlation function
\begin{equation}
\langle\xi_i(t)\xi_j(t')\rangle = \delta_{ij}\delta(t-t'),
\end{equation}
in the amplitude term $T$ is the temperature measured in units where $k_B=1$.
When the trap is applied in a gas phase the friction coefficient can be significantly modified and one can go from the underdamped to overdamped regimes with the same experimental set up if one can control the pressure in the trapping cell.

A semi-phenomenological relationship the  friction coefficient in air as a function of the pressure is given by \cite{beresnev1990,Li2013}
\begin{equation}
\gamma = 6\pi\eta_{atm} R_p \frac{0.619}{0.619+Kn} (1+c_K)
\end{equation}
where in the first,  Stokes drag-like, term $R_p$ is the radius of the particle and $\eta_{\rm atm}$ the viscosity coefficient of air at atmospheric pressure. In addition, $Kn$ is the Knudsen number,  accounting from the deviation from continuum behavior, $Kn = \lambda_{\rm mfp}/R_p$ where $\lambda_{\rm mfp}$ is the mean free path of air molecules and is given by
$\lambda_{\rm mfp}= (68{\rm nm}) P_{\rm atm}/P$, where $P$ is the applied pressure in the 
trap and $P_{\rm atm}$ is the atmospheric pressure. Finally  $c_K=(0.31 Kn) / (0.785+1.152Kn+Kn^2)$. We thus see that the value of the friction coefficient can be tuned by modifying the pressure of the trap.
\subsection{Model of the optical trap}
In the experiment described in the accompanying paper \cite{letter}, the wave length of the laser is  $\lambda =1064 {\rm nm}$, while the radius of the trapped (fused silica) particle is $R_p= 68 {\rm nm}$.
As $R_p\ll \lambda$,  we are in the Rayleigh regime, where the particle can be treated as a point dipole.

Here for completeness we review the model of the optical trap used in \cite{quid2013} as well as reviewing the underlying physics giving rise to both conservative and non-conservative forces in the dipole approximation \cite{hecht2006}.
The Lorentz force acting on an electric dipole ${\bf p}$ is given by
\begin{equation}
{\bf F} = ({\bf p}\cdot\nabla){\bf E} + \frac{d{\bf p}}{dt}\times {\bf B}.
\end{equation}
The response of the dipole to the applied electric field is taken to be linear
\begin{equation}
{\bf p}(t) =\int_{-\infty}^{\infty} dt'\ \alpha(t-t'){\bf E}({\bf r},t).
\end{equation}
For a mono-chromatic electromagnetic field of angular frequency $\omega$ polarized in the $x$ direction we can write
\begin{equation}
{\bf E}({\bf r},t) =  E_0({\bf r})\cos(\phi({\bf r})-\omega t){\bf e}_x,
\end{equation}
where ${\bf e}_x$ denotes the unit vector in the direction ${\bf x}$ and $\phi({\bf r})$ is the phase. Using the Maxwell equation $\partial {\bf B}/\partial t = -\nabla\times {\bf E}$ we can write
\begin{equation}
{\bf B} =\frac{1}{\omega}\nabla\times [E_0({\bf r})\sin(\phi({\bf r})-\omega t){\bf e}_x]
\end{equation}

Using linear response we can write
\begin{eqnarray}
{\bf p}(t) &=&{\rm Re}\  \exp(-i\omega t)\int_{-\infty}^{\infty} dt'\ \alpha(t-t'){\bf E}_0({\bf r})\exp(
i\phi({\bf r})-i\omega(t'-t)) =\nonumber \\ {\rm Re}\ {\bf E}({\bf r},t) \alpha (\omega)
&=& [\alpha'(\omega)\cos(\phi({\bf r})-\omega t)-\alpha''(\omega)\sin(\phi({\bf r})-\omega t) ]{\bf e}_x,
\end{eqnarray}
so $\alpha(\omega)$ is the polarizability at frequency $\omega$ and $\alpha'$ and $\alpha''$ denote the real (dispersive) and imaginary (dissipative)  parts respectively.
Now evaluating the force and replacing the terms $\cos^2(\phi({\bf r})-\omega t)$ and 
$\sin^2(\phi({\bf r})-\omega t)$ by their temporal average $1/2$ and cross terms 
$\cos(\phi({\bf r})-\omega t)$ and $\sin(\phi({\bf r})-\omega t)$ by their temporal average $0$ we obtain
\begin{equation}
{\bf F} = \frac{1}{4}\alpha'\nabla I({\bf r}) + \frac{1}{2}\alpha''I({\bf r}) \nabla \phi({\bf r}), 
\end{equation}
where $I({\bf r}) = |{\bf E}({\bf r})|^2$ is the field intensity. The first term above is clearly a 
conservative force generated by the potential
\begin{equation}
V({\bf r}) = \frac{\alpha'}{4}I({\bf r}).
\end{equation}
The second term contains in general a non-conservative component. Its underlying origin 
comes from the interaction of the magnetic field with the rate of change of the dipole moment. 

The electric field produced by the laser is approximated by the Gaussian form
\begin{equation}
{\bf E}({\bf r}) = E_0 \left[1+\frac{z^2}{z^2_0}\right]^{-\frac{1}{2}}\exp\left[-\frac{x^2}{w^2_x(z)}-\frac{y^2}{w^2_y(z)}+i\phi({\bf r})\right]{\bf e_x},\label{e0}
\end{equation}
where the field is polarized in the direction ${\bf e}_x$ and for  $\alpha = x,\ y$ we have
\begin{equation}
w_\alpha(z) = w_\alpha\left[1+\frac{z^2}{z^2_0}\right]^{\frac{1}{2}},
\end{equation}
is the $z$ dependent beam radius. The phase is given by
\begin{equation}
\phi({\bf r}) = kz -{\rm arctan}\left(\frac{z}{z_0}\right) +\frac{k (x^2+y^2)}{2z(1+\frac{z_0^2}{z^2})},\label{ephi}
\end{equation}
where $k=2\pi/\lambda$ is the wave vector. The parameter $z_0$ is the Rayleigh range and is normally used as a fitting parameter, however it is predicted to be given by
$z_0\sim \pi w_\alpha^2/ \lambda$. Finally, the particle is assumed to be spherical and the 
polarizability $\alpha$ is taken to be \cite{hecht2006}
\begin{equation}
\alpha = \frac{\alpha_0}{1-i\frac{k^3\alpha_0}{6\pi\varepsilon_0}},
\end{equation}
where $\alpha_0$ is the zero-frequency polarizability given by
\begin{equation}
\alpha_0 = 4\pi\epsilon_0 R_p^3\frac{\epsilon-1}{\epsilon+2},
\end{equation}
and $\epsilon_0$ is the dielectric constant of the vaccum while $\epsilon$ is the relative
dielectric constant of the particle.
The form of the electric field given by Eq. (\ref{e0}) arises from the paraxial approximation to Maxwell's equation for the electric field in the vacuum, and in this case the corresponding Helmholtz equation.
The paraxial approximation assumes that for the whole electric field function denoted by $E$ (polarized in the direction $x$) one has that $|\partial^2E/\partial z^2| \ll |k\partial E/\partial z|$.
Inspection of Eqs. (\ref{e0}) and (\ref{ephi}) show that the paraxial approximation is valid when $z/z^2_0 \ll k$, and one can expand the force perturbatively for small $z$ when $z/z_0\ll 1$. The first inequality implies that we must have $z\ll 2\pi^3\omega^4_\alpha/\lambda^3$. 

In the accompanying letter \cite{letter} the laser waist parameters 
are estimated in the high pressure regime where both the effects of trap nonlinearities and the scattering forces are weak. 
It is found that $w_x = 0.915 \mu m$, $w_y = 1.034 \mu m$ and 
$z_0 = w_z = 2.966\mu m$ (in the notation of \cite{letter}) and so we must have $z\ll  0.29 \ {\rm \mu m}$. In addition, one may also Taylor expand the forces when $x/w_x$ and $y/w_z$ are small, experimental results show that these two values are both of the order of $0.1$, so while non-linear effects are weak they are not completely negligible. 

Under the above assumptions, to lowest order, the potential part of the force is generated by a quadratic potential 
\begin{equation}
V({\bf r}) = \kappa_x \frac{x^2}{2} + \kappa_y \frac{y^2}{2}+ \kappa_z\label{harmonic}
\frac{ z^2}{2}
\end{equation}
 with spring constants given by $\kappa_x= E_0^2\alpha'/w_x^2$, $\kappa_y=E_0^2 \alpha'/w_y^2$ and 
$\kappa_z= E_0^2\alpha'/2z^2_0$. 

The phase factor $\phi$ is given, expanding to first and next order of the particle displacement, by
\begin{equation}
\phi({\bf r}) = z\left(k-\frac{1}{z_0}\right) + \frac{z}{z_0^2}\left[\frac{k}{2}(x^2+y^2)
+\frac{z^2}{3z_0}\right].
\end{equation}

As mentioned above, the  term $z_0$ is often treated as a fitting parameter as the solution Eq. (\ref{e0}) is an approximate solution. However an alternative physical approach is to insist that the approximated solution obeys Maxwell's equations exactly near the center of the trap. 
Expanding the solution for small $x$, $y$ and $z$ up to cubic order we find  that 
\begin{equation}
E_x({\bf r}) \approx E_0 \left\{ 1- \frac{x^2}{w_x^2} - \frac{y^2}{w_y^2}- \frac{z^2}{2}\left(\frac{2}{z_0^2} + k^2 -\frac{2k}{z_0}\right) + i\left[z\left(k-\frac{1}{z_0}\right) + \frac{kz(x^2+y^2)}{2z_0^2}
+ \frac{z^3}{3z_0^3}-\frac{z^3}{6}\left(k-\frac{1}{z_0}\right)^3\right]\right\}
\end{equation}
Now using the Helmholtz equation $\nabla^2 E_x + k^2 E_x=0$ near the origin we find the ${\cal O}(0)$ equation, coming from the real term, 
\begin{equation}
\frac{1}{z^2_0}- \frac{k}{z_0} + \frac{1}{w_x^2} + \frac{1}{w^2_y} =0,\label{z0}
\end{equation}
giving an exact relation between $z_0$ and the two waists $w_x$ and $w_y$ in terms of the wave vector $k$. In the case where $w_x=w_y=w$ and when $k\gg 1/z_0$ (so we assume that  the paraxial approximation is valid) we obtain  the standard formula $z_0
 =w^2k/2 = \pi w^2/\lambda$.  However this formula is only valid if $ 2\pi^2 w^2/\lambda^2 \gg 1$. 
 Taking the root of Eq. (\ref{z0}) compatible with the standard formula we find
 \begin{equation}
 z_0 =\frac{2}{k\left[ 1- \left(1-\frac{4}{k^2 w_x^2}-\frac{4}{k^2 w_y^2}\right)^{\frac{1}{2}}\right]},
 \end{equation}
 or in terms of the wavelength $\lambda$ we have
 \begin{equation}
 z_0 = \frac{\lambda}{\pi}\left[1-\left(1-\frac{\lambda^2}{\pi^2 w_x^2} -\frac{\lambda^2}{\pi^2 w_y^2} \right)^{\frac{1}{2}}\right]^{-1}.
 \end{equation} 
 Using this to determine $z_0$, with the values of $w_x$ and 
$w_y$ determined above yields $z_0 =2.781 \mu m$, which is 
different from the fully fitted value of $z_0$ by about $6\%$. This estimate thus provides a useful starting point to determine the trap parameters in the fitting procedure. 

 The scattering part of the force is given to leading, quadratic, order by
 \begin{equation}
 {\bf F}_{\rm scat} =\frac{\alpha''}{2} E_0^2\left\{ {\bf e}_x \frac{kxz}{z_0^2} +  {\bf e}_y \frac{kyz}{z_0^2} + {\bf e}_z\left[(k-\frac{1}{z_0})\left(1-\frac{2x^2}{w_x^2}-\frac{2y^2}{w_y^2}-\frac{z^2}{z_0^2}\right)
 + k\frac{x^2+y^2}{2 z^2_0} +\frac{z^2}{z_0^3}\right]\right\},
 \end{equation}
 which can be written in terms of the spring constant $\kappa_z$ as
 \begin{equation}
 {\bf F}_{\rm scat} =\frac{\alpha''}{\alpha'}\kappa_z \left\{ {\bf e}_x kxz +  {\bf e}_y kyz + {\bf e}_z\left[z_0(z_0k-1)\left(1-\frac{2x^2}{w_x^2}-\frac{2y^2}{w_y^2}-\frac{z^2}{z_0^2}\right)
 + k\frac{x^2+y^2}{2 } +\frac{z^2}{z_0}\right]\right\},
 \end{equation}
 which gives
 \begin{equation}
 {\bf F}_{\rm scat}= \frac{\alpha''}{\alpha'}\kappa_z \left\{ {\bf e}_x kxz +  {\bf e}_y kyz + {\bf e}_z\left[z_0(z_0k-1)-{x^2}\left(\frac{2z_0(z_0k-1)}{w_x^2} -\frac{k}{2}\right) - {y^2}\left(\frac{2z_0(z_0k-1)}{w_y^2} -\frac{k}{2}\right) -z^2 \left(k -\frac{2}{z_0}\right)\right]\right\}.\label{fscatl}
 \end{equation}
 The lowest order approximation that takes into account both conservative/potential  and scattering forces is thus the expressions for ${\bf F}_p=-\nabla V$ (${\cal O}(1)$) and ${\bf F}_{\rm scat}$ (${\cal O}(0) + {\cal O}(2)$). The next order corrections are ${\cal O}(3)$ terms to ${\bf F}_p$ coming from anharmonic contributions.
 
 Using the form Eq. (\ref{e0}) in the Helmholtz equation without making any approximations yields
 \begin{equation}
 \nabla^2 E_x -[\nabla\phi]^2 E_x + k^2 E_x + i\left(2\nabla  E_x \cdot \nabla \phi + E_x \nabla^2\phi\right)=0,\label{eqh}
 \end{equation}
 The real and imaginary parts of the above give two equations. The scattering force is given by
 \begin{equation}
 {\bf F}_{\rm scat}=\frac{\alpha''}{2}I\nabla \phi = \frac{\alpha''}{2} E_x^2\nabla \phi 
 \end{equation}
 and taking the divergence of this yields
 \begin{equation}
 \nabla\cdot {\bf F}_{\rm scat} = \frac{\alpha''}{2}E_x\left(2\nabla  E_x \cdot \nabla \phi + E \nabla^2\phi\right)=0,
 \end{equation}
 from Eq. (\ref{eqh}). However the result Eq. (\ref{fscatl}) gives $\nabla\cdot {\bf F}_{\rm scat}
 = 4z\alpha''\kappa_z/\alpha'z_0$. In the limit $z_0\to \infty$, where the paraxial approximation becomes exact we find that this term is zero, but in general it is not.  This actually mean that the very last term in Eq. (\ref{fscatl}) should not be there. This can be regarded as being due to an error in the phase term.
 
 Taking this into account we can naturally divide the scattering force into two components
 \begin{equation}
 {\bf F}_{\rm scat}= {\bf F}_{\rm scat,1} + {\bf F}_{\rm scat,2},
 \end{equation}
 where 
 \begin{equation}
 {\bf F}_{\rm scat,1}  = \frac{\alpha''}{\alpha'}\kappa_z z_0(z_0k-1)\left[1-\frac{2x^2}{w_x^2} - \frac{2y^2}{w_y^2} \right] {\bf e}_z.\label{fscat1}
 \end{equation}
 and 
 \begin{equation}
 {\bf F}_{\rm scat,2}= \frac{\alpha''}{\alpha'}k\kappa_z \left[ {\bf e}_x xz +  {\bf e}_y yz - {\bf e}_zz^2\right] \label{fscat2}.
 \end{equation}
 For the values of $z_0$ obtained here the dominant scattering force is ${\bf F}_{\rm scat,1}$, 
 however as $z$ scales as $z_0$ when just the harmonic confinement is taken into account it is not evident that the last term in the direction ${\bf e}_z$ can be neglected.
 However we find  that numerically this turns out to be the case.
 
\section{Perturbative calculation of the non-equilibrium steady state probability density function and current}

In the standard paradigm of statistical physics, Brownian particles are
subjected to forces generated by potentials and this means that the underlying steady state distribution is given by the Gibbs-Boltzmann distribution. The crucial point here is that the steady state in the presence of a nonconservative force has an associated current. We note that Brownian particles subject to linear potentials in unbounded 
systems also have currents, however because the system is unbounded the probability distribution spreads with time and can be 
associated with effective transport coefficients \cite{dea2007}. However in periodic systems the particle position modulo the period of the system does have a steady state which carries a current, and this system can be analysed for overdamped Brownian particles \cite{reim2001}. In
general very few results exist on the steady states of systems with nonconservative forces or in the presence of driving, even for trapped mono-particle systems where only one body forces are present. Here we will resort to the use of perturbation theory, which can be justified when the nonconservative component of the force is weak in comparison to the conservative component. Indeed, perturbation theory was the approach employed in \cite{grier2015} to analyse a trapped overdamped Brownian particle.

We will consider  systems which are driven by underlying white noise, and thus include both underdamped and overdamped Brownian motion. The probability distribution of the particle's position in phase space is thus described by the Fokker-Planck equation
\begin{equation}
\frac{\partial P({\bf y})}{\partial t} = -H P
\end{equation}
where ${\bf y}$ denote the phase-space coordinates and $H$ is the forward Fokker-Planck operator. Before perturbation by a non-conservative force we assume that the system obeys detailed balance and has an underlying Gibbs-Boltzmann distribution denoted by $P_0$, obeying
\begin{equation}
HP_0 = 0.
\end{equation}

Now we consider what happens when the system is perturbed by an additional force
which is in terms of the Fokker-Planck operator denoted by the perturbation $\Delta H$. The question we address  is what is the new steady state distribution $P_s$ defined by
\begin{equation}
[H+\Delta H]P_s = 0.
\end{equation}
To first order in perturbation theory we write $P_s= P_0 +\Delta P_0$ where $\Delta P_0$ is given by
\begin{equation}
H\Delta P_0 = -\Delta H P_0.
\end{equation}
This now has the formal solution
\begin{equation}
\Delta P_0({\bf y}) = -\int d{\bf y}' H^{-1}({\bf y},{\bf y}') [\Delta H P_0]({\bf y}').
\end{equation}
Here $H^{-1}$ is the pseudo-Green's function for the operator $H$ as the solution 
$\Delta H P_0$ must obey the normalization 
\begin{equation}
\int d{\bf y} \Delta P_0({\bf y}) = 0.
\end{equation}
The transition probability $P({\bf y}|{\bf y}';t)$ can formally be written as
\begin{equation}
P({\bf y}|{\bf y}';t) = \exp(-tH)({\bf y},{\bf y}')
\end{equation}
and the pseudo-Green's function thus as
\begin{equation}
H^{-1}({\bf y},{\bf y}') = \int_0^\infty dt [P({\bf y}|{\bf y}';t) - P_0({\bf y})].
\end{equation}
This yields 
\begin{equation}
\Delta P_0({\bf y}) = -\int_0^\infty dt\int d{\bf y}' [P({\bf y}|{\bf y}';t) - P_0({\bf y})] [\Delta H P_0]({\bf y}'). 
\end{equation}
Now using the time reversal symmetry of the equilibrium state 
\begin{equation}
P({\bf y}|{\bf y}';t)=\frac{P({\bf y}'^*|{\bf y}^*;t)P_0({\bf y}^*)}{P_0({\bf y}')},
\end{equation}
where $^*$ indicates time reversal of the coordinates, this means that coordinates with an odd number of temporal 
derivatives change sign. Taking $t=0$ we see that time reversal symmetry also implies that $P_0({\bf y}^*)=P_0({\bf y})$. For overdamped Brownian particles this has no effect as the position vector obeys
${\bf X}={\bf X}^*$. However if ${\bf Y}$ represents the position and velocity ${\bf Y}=({\bf X},{\bf V})$ we have
${\bf Y}^* = ({\bf X},-{\bf V})$. We thus find
\begin{equation}
\Delta P_0({\bf y}) = -P_0({\bf y})\int_0^\infty dt\int d{\bf y}' [P({\bf y}'^*|{\bf y}^*;t) - P_0({\bf y'})] \frac{[\Delta H P_0]({\bf y}')}{P_0({\bf y'})}. 
\end{equation}
If the stochastic process represented by the Fokker-Planck equation is denoted by ${\bf Y}_t$ and we use the notation 
\begin{equation}
\langle f({\bf Y}_t)\rangle_{\bf y} = \int d{\bf y'} P({\bf y}'|{\bf y};t)f({\bf y}'),
\end{equation}
i.e. the average value of $f({\bf Y}_t)$ at time $t$ given that ${\bf Y}_0 ={\bf y}$, and
\begin{equation}
\langle f({\bf Y}_t)\rangle_{0} = \int d{\bf y'} P_0({\bf y})f({\bf y}'),
\end{equation}
the equilibrium value of $f({\bf Y})$.
We thus change the integration variable to write
\begin{equation}
\Delta P_0({\bf y}) = -P_0({\bf y})\int_0^\infty dt\int d{\bf y}' [P({\bf y}'|{\bf y}^*;t) - P_0({\bf y'})] \frac{[\Delta H P_0]({\bf y}'^*)}{P_0({\bf y'}^*)}. 
\end{equation}
where we have used ${\bf y}^{**}={\bf y}$. This then yields
\begin{equation}
\Delta P_0({\bf y}) = -P_0({\bf y})\int_0^\infty dt\left[\left\langle  \frac{[\Delta H P_0]({\bf Y}^*_t)}{P_0({\bf Y}^*_t)}\right\rangle_{{\bf y}^*} - \left\langle\frac{[\Delta H P_0]({\bf Y}^*)}{P_0({\bf Y}^*)}\right\rangle_{0}\right].
\end{equation}
The change in the equilibrium distribution at the point ${\bf y}$ in phase space is thus given by a Kubo-type formula \cite{kubo}. This result is general and applies to any system described by a Fokker-Planck equation.
\subsection{Overdamped Brownian motion}
Here we consider a process characterized solely by the position ${\bf X}$, that is to say the overdamped regimes. Here we assume that the unperturbed 
Fokker Planck operator, defined via its action on a test function $f$, is given by
\begin{equation}
Hf = -\nabla \cdot D\left[\nabla f + \beta f\nabla V\right],
\end{equation}
{\em i.e.} it describes a  particle with diffusivity $D$ in a potential $V$ at temperature $T$
(where for convenience we use units where $k_B=1$, the results for $T$ measured in Kelvin are thus obtained by replacing $T$ by $k_BT$ in all the formulas which follow) and $\beta =1/T$.

Here the steady state is given by the Gibbs-Boltzmann equilibrium distribution
\begin{equation}
P_0({\bf x}) = \frac{1}{Z}\exp\left(-\beta V({\bf x})\right).
\end{equation}
The perturbation of the Fokker-Planck operator due to the presence of an extra force ${\bf F}$ is given by
\begin{equation}
\Delta H f = \nabla \cdot D\beta {\bf F} f.
\end{equation}

From this we find 
\begin{equation}
\frac{[\Delta H P_0]({\bf x})}{P_0({\bf x})}= -\beta^2 D\nabla V({\bf x}) \cdot {\bf F}({\bf x}) + 
\beta D\nabla\cdot {\bf F}.
\end{equation}
This now gives
\begin{equation}
\Delta P_0({\bf x}) = P_0({\bf x})\beta\int_0^\infty dt\left[\langle \beta D\nabla V({\bf X}_t) \cdot {\bf F}({\bf X}_t) - 
D\nabla\cdot {\bf F}({\bf X}_t) \rangle_{\bf x} - \langle \beta D\nabla V({\bf X}) \cdot {\bf F}({\bf X}) +
D\nabla\cdot {\bf F}({\bf X})\rangle_{0}\right].\label{bmgen}
\end{equation}
The case where the force is derived from a potential,  ${\bf F}=-\nabla U$, should of course be simple as we know the new equilibrium measure should be
\begin{equation}
P_s({\bf x}) = \frac{\exp\left(-\beta V({\bf x})-\beta U({\bf x})\right)}{\int d{\bf x} \exp\left(-\beta V({\bf x})-\beta U({\bf x})\right)},
\end{equation}
and for small $U$ should find
\begin{equation}
P_s({\bf x}) = P_0({\bf x})\left( 1-\beta U({\bf x}) +\beta \langle U({\bf X})\rangle_0\right).\label{gp}
\end{equation}
It is illuminating, as well as a useful check,  to  obtain this from the general formalism derived previously. To do this, we use the following result of stochastic calculus \cite{osk} for an arbitrary function $U$:
\begin{equation}
\left\langle \frac{ dU({\bf X_t})}{dt}\right\rangle = - D\beta \nabla U({\bf X_t})\cdot \nabla V({\bf X_t})
+ D\nabla^2 U({\bf X}_t).
\end{equation}
Using this and the relation ${\bf F}=-\nabla U$ in Eq. (\ref{bmgen}) now gives
\begin{equation}
\Delta P_0({\bf x}) = P_0({\bf y})\beta\int_0^\infty dt \left[\left\langle \frac{ dU({\bf X_t})}{dt}\right\rangle_{\bf x}- \left\langle\frac{ dU({\bf X})}{dt}\right\rangle_{0}\right].\label{int1}
\end{equation}
Clearly we have 
\begin{equation}
\left\langle\frac{ dU({\bf X})}{dt}\right\rangle_{0} =0,
\end{equation}
and also we have by definition that
\begin{equation}
\lim_{t\to \infty }\langle  U({\bf X}_t)\rangle _{\bf x}  = \langle  U({\bf X})\rangle _{0}.
\end{equation}
The time integral in Eq. (\ref{int1}) can now be simply carried out to yield
\begin{equation}
\Delta P_0({\bf x}) = P_0({\bf x})\left[ -\beta U({\bf x}) + \beta\langle  U({\bf X})\rangle _{0}\right],
\end{equation}
in agreement with Eq. (\ref{gp}). 

In general the perturbing force can be decomposed into
a conservative and nonconservative part via the Helmholtz decomposition where one
writes
\begin{equation}
{\bf F} = -\nabla U + \nabla\times {\bf A},
\end{equation}
where $U$ is determined from the equation
\begin{equation}
\nabla^2 U = -\nabla\cdot{\bf F},
\end{equation}
and the remaining term can be similarly evaluated to write
\begin{equation}
{\bf F} = -\nabla U + {\bf F}_n
\end{equation}
where the nonconservative part of the force ${\bf F}_n$ is given by 
\begin{equation}
{\bf F}_n = {\bf F} +\nabla U, 
\end{equation}
however, the solution for $U$ must be chosen so that $\nabla\cdot {\bf F}_n =0$. Using this decomposition then yields
\begin{equation}
\Delta P_0({\bf x}) = \beta P_0({\bf x})\left[ - U({\bf x}) + \langle  U({\bf X})\rangle_0 +\int_0^\infty dt\left[\langle \beta D\nabla V({\bf X}_t) \cdot {\bf F}_n({\bf X}_t)  \rangle_{\bf x} - \langle \beta D\nabla V({\bf X}) \cdot {\bf F}_n({\bf X}) 
\rangle_{0}\right]\right].
\end{equation}
The modification of the steady state probability density function by the non-conservative force is thus seen to be related to the work done by the non-conservative force in the direction of the original potential $V$. 
This formula is not particularly illuminating from a physical point of view for overdamped Brownian motion, when we study the case of general damping we will find a much more explicit interpretation.
\subsubsection{Application to a model for an optical trap}

Here we study the model proposed by Grier et al \cite{grier2015}, where
\begin{equation}
V({\bf x}) = \frac{\kappa}{2}(x^2+y^2 + \eta z^2),\label{V1}
\end{equation}
 so this represents an anisotropic harmonic trap unless $\eta=1$. We have seen that the potential part of the scattering force in Eq. (\ref{harmonic}) in general also has $\kappa_x\neq \kappa_y$ as the difference between $\kappa_x$ and $\kappa_y$ is numerically small we carry out the analysis for this case to simplify the notation and the 
 following analysis.
 In \cite{grier2015} the nonconservative component of the force generated by the optical trap is taken to be
\begin{equation}
{\bf F} = \varepsilon\kappa a\left(1-\frac{x^2+y^2}{a^2}\right){\bf e}_z,\label{F1}
\end{equation}
$a$ is proportional to the waist $w_x\approx w_y$ (we use the same notation as \cite{grier2015} for ease of comparison). The part of the scattering force used is thus
${\bf F}_{\rm scat,1}$, expressed in Eq.~(\ref{fscat1}) which is that which is numerically the most important. However the nature of our perturbative computation means that the corrections to $\Delta P_0$ from  each component of the non-conservative force is additive, and so we can simply compute the perturbations due to each component and then sum them. 
 The perturbing parameter $\varepsilon$ is dimensionless and we will compute the 
modified steady state probability density function to first order in $\varepsilon$. 

The decomposition of the perturbation in terms of a conservative and nonconservative contribution now yields the potential
\begin{equation}
U({\bf x}) = -\varepsilon \kappa a z
\end{equation}
and the nonconservative component of the force
\begin{equation}
{\bf F}_n = -\varepsilon\kappa\frac{x^2+y^2}{a}\,{\bf e}_z.\label{ncz}
\end{equation}
Using this we find
\begin{equation}
\nabla V\cdot {\bf F}_n = -\varepsilon\kappa^2  \eta z \frac{x^2+y^2}{a}.
\end{equation}
The evaluations of the expectation values required to compute the renormalization of the steady state probability distribution can be carried out by explicitly solving the Langevin equation for the unperturbed process, the corresponding equations are
\begin{eqnarray}
\dot x(t) &=& -\beta D\kappa x(t) + \sqrt{2D}\xi_x(t), \\
\dot y(t) &=& -\beta D\kappa y(t) + \sqrt{2D}\xi_y(t), \\
\dot z(t) &=& -\beta D\kappa\eta z(t) + \sqrt{2D}\xi_z(t), 
\end{eqnarray}
where $\xi_x(t)$, $\xi_y(t)$ and $\xi_z(t)$ are independent Gaussian white noises. The above equations can be solved in terms of the particle positions at $t=0$, denoted by $x$, $y$ and $z$ to give
\begin{eqnarray}
x(t) &=&  \exp(-\beta D\kappa t)x +  \sqrt{2D} \exp(-\beta D\kappa t)\int_0^t ds \exp(\beta D\kappa s)   \xi_x(s),\\
y(t) &=&  \exp(-\beta D\kappa t)y +  \sqrt{2D} \exp(-\beta D\kappa t)\int_0^t ds \exp(\beta D\kappa s)   \xi_y(s) ,    \\
z(t) &=&     \exp(-\beta D\kappa\eta t)z +  \sqrt{2D} \exp(-\beta D\kappa\eta t)\int_0^t ds \exp(\beta D\kappa\eta s)   \xi_z(s) . 
\end{eqnarray}
Using these solutions we find
\begin{equation}
\langle \nabla V\cdot {\bf F}_n \rangle_{\bf x} = -\varepsilon\kappa^2\eta \exp(-\beta D\kappa\eta t)z \frac{(x^2+y^2)\exp(-2\beta D\kappa t) + \frac{2}{\kappa\beta}(1-\exp(-2\beta D\kappa t)}{a}.
\end{equation}
and we also see that
\begin{equation}
\langle \nabla V\cdot {\bf F}_n \rangle_{0}=0 .
\end{equation}
Putting all this together (and noting that $\langle U({\bf X})\rangle_0=0$) we obtain
\begin{equation}
\label{gammaoverdamp}
\frac{\Delta P_0({\bf x})}{P_0({\bf x})} = -\beta\Delta\phi = \frac{\varepsilon z}{a}\left[\beta\kappa a^2-2 + \frac{\eta}{\eta+2}(2-\beta\kappa [x^2+y^2])\right],
\end{equation}
where $\Delta \phi$ is interpreted as the renormalization of the trapping potential. This agrees with the result found in \cite{grier2015} for the same trapping potential and nonconservative perturbation considered here.
\subsubsection{Alternative formulation for overdamped particles}
The probabilistically based solution for the perturbed steady state distribution gives us an interpretation of the 
perturbation of the steady state in terms of work done by the perturbing forces. However in the case where the 
unperturbed solution is Gaussian we see that the solution is particularly simple, the solution having a polynomial form 
for a polynomial perturbing force. Given the structure of the solution found we make the same ansatz as in \cite{grier2015}
\begin{equation}
\Delta P_0({\bf x}) = P_0({\bf x})K({\bf x}),
\end{equation}
this yields
\begin{equation}
\nabla^2 K({\bf x}) -\beta\nabla V({\bf x}) \cdot \nabla K({\bf x}) = \beta \nabla \cdot {\bf F}({\bf x}) -\beta^2 \nabla V({\bf x})\cdot {\bf F}({\bf x}).
\end{equation}
For the case studied above as defined by the potential in Eq. (\ref{V1}) and the force in Eq. (\ref{F1}) we  find
\begin{equation}
\nabla^2 K -\beta \kappa  x\frac{\partial K}{\partial x} -\beta \kappa y \frac{\partial K}{\partial y}
- \beta \kappa \eta z \frac{\partial K}{\partial z} = -\varepsilon\beta^2\kappa^2a \eta z\left(1-\frac{x^2+y^2}{a^2}\right).
\end{equation}
Inspection of the probabilistic representation of the solution, and from the symmetry of the problem in the $(x,y)$ plane, now tells us that the solution must have the form
\begin{equation}
K({\bf x}) = z\left[A+ B(x^2+y^2)\right],
\end{equation}
this yields 
\begin{equation}
4B - 2B\beta\kappa(x^2+y^2) -\beta\kappa\eta \left[ A + B(x^2+y^2)\right] = -\varepsilon\beta^2\kappa^2a \eta \left(1-\frac{x^2+y^2}{a^2}\right),
\end{equation}
and comparing the coefficients of the polynomials now gives
\begin{eqnarray}
A &=& \frac{\varepsilon}{(\eta + 2)a}\left[-4 +\beta \kappa a^2(\eta+2)\right],\\
B &=& -\frac{\varepsilon\beta\kappa\eta}{(\eta+2)a},
\end{eqnarray}
in agreement with the earlier calculation leading to Eq. (\ref{gammaoverdamp}) and thus the result given in \cite{grier2015}.

Here we also consider the perturbation to a force of type ${\bf F}_{\rm scat,2}$ derived in Eq.~(\ref{fscat2}) which we write as
\begin{equation}
 {\bf F}'={\bf F}'_n=  \frac{\varepsilon'\kappa}{a} \left[ {\bf e}_x xz +  {\bf e}_y yz - {\bf e}_z  z^2\right] \label{fprime}
\end{equation}
as it has no potential part. Here we find
\begin{equation}
\nabla^2 K' -\beta \kappa  x\frac{\partial K'}{\partial x} -\beta \kappa y \frac{\partial K'}{\partial y}
- \beta \kappa \eta z \frac{\partial K'}{\partial z} =-\frac{\varepsilon'\beta^2\kappa^2}{a}z(x^2+y^2-\eta z^2),\label{F2}
\end{equation}
where the $'$ on $K$ denotes that it is the form of $K$ for the perturbation ${\bf F}'$. Here we see that the solution must be of the form
\begin{equation}
K'({\bf x}) = z\left[A' + B'(x^2+y^2) + J'z^2\right].
\end{equation}
This gives
\begin{equation}
(6J' + 4B')z -2\beta\kappa B' zx^2 -2\beta\kappa B' zy ^2 -\beta\kappa\eta z(A'+3J'z^2 + B'x^2 + B'y^2) = -\frac{\varepsilon'\beta^2\kappa^2}{a}z(x^2+y^2-\eta z^2),
\end{equation}
and thus
\begin{eqnarray}
A'&=& -\frac{2\varepsilon'}{a(2+\eta)}\\
B'&=&  \frac{\varepsilon'\beta\kappa}{a(2+\eta)} \\
J'&=& -\frac{\varepsilon'\beta\kappa}{3a}.
\end{eqnarray}
This therefore gives a contribution to the steady state distribution
\begin{equation}
\frac{\Delta P_s({\bf x})}{P_0({\bf x})}=\frac{\beta \kappa \varepsilon' z}{a}\left[\frac{1}{2+\eta}\left(x^2+y^2-\frac{2}{\beta\kappa}\right) -\frac{1}{3} z^2\right]\label{pertoverprime}
\end{equation}
\subsection{Underdamped Brownian particles}
Here the phase space corresponds to ${\bf Y}= ({\bf X},{\bf V})$ where ${\bf X}$ and ${\bf V}$ denote respectively the particle position and its velocity. Here the Fokker-Planck operator is given by
\begin{equation}
\label{equnderdamped}
Hf({\bf x},{\bf v}) =  -\nabla_{\bf v}\cdot\left[\frac{T\gamma}{m^2}\nabla_{\bf v} f + \frac{\gamma}{m}{\bf v} f
-\frac{1}{m}{\bf F}_t({\bf x})f\right] +\nabla_{\bf x} \cdot\left[{\bf v} f\right].
\end{equation}
where $\gamma$ is the friction coefficient, $m$ is the particle mass, and ${\bf F}_t$ is the total force on the particle 
which is assumed to depend only on the position ${\bf x}$. For the unperturbed system we take ${\bf F}_t({\bf x})=-\nabla V({\bf x})$. The equilibrium distribution is then given by
\begin{equation}
\label{P0underdamped}
P_0({\bf x},{\bf v}) = \frac{1}{Z}\exp\left[-\frac{1}{2}\beta m{\bf v}^2 -\beta V({\bf x})\right] 
\end{equation}

In the perturbed system we denote the additional force simply by ${\bf F}({\bf x})$. This yields
\begin{equation}
\frac{\Delta H P_0({\bf x},{\bf v})}{P_0({\bf x},{\bf v})}= -\beta {\bf v}\cdot {\bf F}({\bf x}),
\end{equation}
and thus 
\begin{equation}
\Delta P_0({\bf x},{\bf v}) = -P_{0}({\bf x},{\bf v})\beta \int_0^\infty  dt\left[ \langle {\bf V}_t\cdot {\bf F}({\bf X}_t)\rangle_{{\bf x},-{\bf v}} - \langle {\bf V}\cdot {\bf F}({\bf X})\rangle_{0}\right],\label{wod}
\end{equation}
where we have used ${\bf V}^*=-{\bf V}$. The result in Eq.~(\ref{wod}) immediately gives us a physical interpretation of the modification of the steady state probability density function. We see that 
\begin{equation}
\int_0^\infty  dt\left[ \langle {\bf V}_t\cdot {\bf F}({\bf X}_t)\rangle_{{\bf x},-{\bf v}} - \langle {\bf V}\cdot {\bf F}({\bf X})\rangle_{0}\right] =\overline  W({\bf x},-{\bf v}) -\overline W_0,
\end{equation}
where $\overline  W({\bf x},-{\bf v})$ is the total work done, after a very long time $t^*$, by the non-conservative force on a particle started at the posiiton ${\bf x}$ but and with initial velocity $-{\bf v}$ , whereas $\overline W_0$ is the total work done by the non-conservative force, at the same late time $t^*$, for a particle whose initial velocity and position are taken from the unperturbed equilibrium probability density function. Both of these total works diverge as $t^*\to \infty$ but their difference is constant.
When the perturbing force is purely conservative, {\em i.e.} ${\bf F}= -\nabla U$ we immediately obtain, much more transparently that in the Brownian case,
\begin{equation}
\Delta P_0({\bf x},{\bf v}) = P_{0}({\bf x},{\bf v})\beta \int_0^\infty  dt\left[ \langle {\bf V}_t\cdot \nabla U({\bf X}_t)\rangle_{{\bf x},-{\bf v}} - \langle {\bf V}\cdot\nabla U ({\bf X})\rangle_{0}\right],\label{kubov}
\end{equation} 
and as ${\bf V}= d{\bf X}/dt$ we find, using the same arguments as for the overdamped case
\begin{equation}
\Delta P_0({\bf x},{\bf v}) = P_{0}({\bf x},{\bf v})\beta\left(\langle U({\bf X})\rangle_0 - U({\bf x})\right),
\end{equation}
thus recovering the corresponding modified Gibbs-Boltzmann distribution. 

\subsection{Application to under generally damped particles in optical trap model}

In principle, for a harmonic trap all the average values in Eq. (\ref{wod}) can be computed. However, as for the overdamped case, the form of the solution suggests writing
\begin{equation}
\Delta P_0({\bf x},{\bf v}) = P_0({\bf x},{\bf v})K({\bf x},{\bf v}).
\end{equation}
This then yields
\begin{equation}
\frac{T\gamma}{m^2}\nabla_{\bf v}^2 K -\frac{\gamma}{m}{\bf v}\cdot\nabla_{\bf v}K -{\bf v}\cdot\nabla_{\bf x} K +\frac{1}{m} \nabla_{\bf x} V({\bf x})\cdot \nabla_{\bf v}K = 
-\beta{\bf F}\cdot {\bf v} .\label{egamma}
\end{equation}
There is a distinct advantage, at least for complex perturbations by non-conservative forces, of using this algebraic formulation as opposed to the probabilistic representation. However the probabilistic representation allows one to  predict the polynomial form of $K$ and only the coefficients of this polynomial need be determined after substituting the correct form of $K$ into Eq. (\ref{egamma}). This is because in the expression $\langle {\bf V}_t\cdot {\bf F}({\bf X}_t)\rangle_{{\bf x},-{\bf v}}$ we can see explicitly what terms in the initial position ${\bf x}$ and velocity ${\bf v}$ must arise by inspecting the form of ${\bf F}$ given in Eq. (\ref{F1}). 

From the symmetry in the $(x,y)$ plane, the solution of Eq.~(\ref{egamma}) must have the form
\begin{equation}
\label{gammaunderdamped}
K({\bf x},{\bf v}) = z [A+B(x^2+y^2)+C(xv_x+yv_y)+D(v_x^2+v_y^2)] + v_z [E+F(x^2+y^2)+G(xv_x+yv_y)+H(v_x^2+v_y^2)].
\end{equation}
After simplification, the coefficients are given by
\begin{eqnarray}
A&=&\beta\varepsilon\kappa a - \frac{8\varepsilon\gamma^2}{S} [6\gamma^2+(8-5\eta)\kappa m], \\
B&=&-\frac{\eta\beta\varepsilon\kappa}{S} [(\eta-4)(\eta-2)(\kappa m)^2+4(2\eta+1)\gamma^2\kappa m+12\gamma^4], \\
C&=&\frac{4\eta\beta\varepsilon\gamma\kappa m }{S}[6\gamma^2+(\eta+2)\kappa m], \\
D&=&-\frac{2\eta\beta\varepsilon\kappa m^2}{S} [6\gamma^2+(4-\eta)\kappa m], \\
E&=&\frac{8\varepsilon\gamma m}{S} [6\gamma^2+(4-\eta)\kappa m], \\
F&=&-\frac{4\beta\varepsilon\gamma\kappa m}{S}[6\gamma^2+(\eta+2)\kappa m] \\
G&=&- \frac{2\beta\varepsilon \kappa m^2}{S}[6(\eta-2)\gamma^2+\eta(\eta-4)\kappa m], \\
H&=&\frac{8\beta\varepsilon \gamma\kappa m^3}{S} (\eta-1),
\end{eqnarray}
where $S=a[\eta\kappa m+2\gamma^2][(\eta-4)^2\kappa m + 6 (2+\eta)\gamma^2]$. Using these expressions we  have thus computed the perturbed probability density function in the full $({\bf x},{\bf v})$ phase space. However the final expressions are rather unwieldy,  we will therefore give explicit expressions for the marginal probability density functions of  ${\bf X}$ and then ${\bf V}$.

Denoting  the marginal steady state probability density function for the position by $\Delta P_0({\bf x}) = \int d{\bf v} K({\bf x},{\bf v}) P_0({\bf x},{\bf v})$, we find from Eq.~(\ref{P0underdamped})-(\ref{gammaunderdamped})
\begin{equation}
\Delta P_0({\bf x}) = P_0({\bf x}) \frac{\varepsilon z}{a} \left[\beta\kappa a^2 -2 +  \eta\left(2- \beta \kappa \rho^2\right) \frac{ (\eta-4)(\eta-2) (\kappa m)^2+4(2\eta+1)\kappa m \gamma^2 +12\gamma^4}{[\eta\kappa m+2\gamma^2][(\eta-4)^2\kappa m + 6 (2+\eta)\gamma^2]} \right],\label{dpgen}
\end{equation}
where $\rho =\sqrt{x^2+y^2}$. In the underdamped  limit $\gamma \rightarrow 0$, we find thus
\begin{equation}
\Delta P_0({\bf x}) =  P_0({\bf x})\frac{\varepsilon z}{a} \left[\beta \kappa a^2 -2 + \frac{\eta-2}{\eta-4}(2-\beta \kappa \rho^2) \right]
\end{equation}
and in the overdamped limit   $m \rightarrow 0$, we recover (for a third time) the result of 
\cite{grier2015} Eq. (\ref{gammaoverdamp}).

Using the modified probability density function in Eq. (\ref{dpgen}) can study various moments of the particle position. The average value of the height of the particle is given by
\begin{equation}
\langle z\rangle  = \frac{\varepsilon a}{\eta} \left(1 - \frac{2}{\beta\kappa a^2} \right),
\end{equation}
while the average velocity in the direction $z$ is given by $\langle V_z\rangle =0$, after some mathematical cancellations which appear remarkable from a mathematical point of view but obvious from the simple physical insight that the particle is trapped!
The above result for $\langle z\rangle$ is thus independent of the details of the dynamics (it is independent of the damping parameters). This surprising result is a consequence of the fact that the average velocity is zero (as obviously must be the case if the particle remains trapped): as the average velocity is zero the mean position along the $z$ axis is given by a force balance equation where the friction term is zero and, as the average thermal force and acceleration are zero,
the average force balance equation depend only on the external forces acting on the particle and is thus independent of the damping parameters.

Here we introduce the underlying  frequency of the harmonic part of the trap $\omega_0^2=\kappa/m$ and the damping rate $\Gamma=\gamma/m$. The quality factor is defined as $Q=\omega_0/\Gamma = \sqrt{\kappa m / \gamma^2}$, it quantifies the degree of damping in the system, large $Q$ corresponding to the underdamped regime and small $Q$ corresponding to the overdamped regime. In terms of these variables, which are often preferred in experimental studies, we find
\begin{equation}
\Delta P_0({\bf x}) = P_0({\bf x})\frac{\varepsilon z}{a} \left[\beta\kappa a^2 -2 +  \eta\left(2- \beta \kappa \rho^2\right) \frac{ (\eta-4)(\eta-2) Q^4+4(2\eta+1)Q^2 +12}{[\eta Q^2+2][(\eta-4)^2 Q^2 + 6 (2+\eta)]} \right].
\end{equation}
Denoting the marignal steady state probability density function for the velocity by $\Delta P_{\rm MB}({\bf v}) = \int d{\bf x} K({\bf x},{\bf v}) P_0({\bf x},{\bf v})$, we find from Eq.~(\ref{P0underdamped})-(\ref{gammaunderdamped})
\begin{equation}
\Delta P_{\rm MB}({\bf v}) = P_{\rm MB}({\bf v}) \frac{8 \beta \kappa \varepsilon v_z}{a} \left(v_\rho^2-\frac{2}{\beta m}\right) \frac{\gamma\kappa m^3 (\eta-1)}{[\eta\kappa m+2\gamma^2][(\eta-4)^2\kappa m + 6 (2+\eta)\gamma^2]} \label{dpv}
\end{equation}
where $v_\rho =\sqrt{v_x^2+v_y^2}$ and
\begin{equation}
P_{\rm MB}({\bf v})=\left(\frac{\beta m}{2\pi}\right)^{\frac{3}{2}}\exp(-\frac{\beta m{\bf v}^2}{2}),
\end{equation}
is the Maxwell-Boltzmann velocity distribution. The marginal probability distributions for $v_x$, $v_y$ and $v_z$ however do not change in first order perturbation theory and are thus given by the Maxwell distribution.

In \cite{grier2015} the steady state current was computed for an overdamped Brownian particle. Here we have currents in both position and velocity spaces. From Eq.~(\ref{equnderdamped}), we see that currents ${\bf J_x}$ and ${\bf J_v}$ can be read off from
\begin{equation}
\frac{\partial P({\bf x},{\bf v},t)}{\partial t} = -HP({\bf x},{\bf v},t) = -\nabla_{\bf x} \cdot {\bf J_x}({\bf x},{\bf v},t) - \nabla_{\bf v} \cdot {\bf J_v}({\bf x},{\bf v},t),\label{fpc}
\end{equation}
where 
\begin{eqnarray}
{\bf J_x}({\bf x},{\bf v},t) &=& {\bf v}P({\bf x},{\bf v},t) \label{currx}\\
{\bf J_v}({\bf x},{\bf v},t) &=& -\frac{T\gamma}{m^2}\nabla_{\bf v} P({\bf x},{\bf v},t) - \frac{\gamma}{m}{\bf v} P({\bf x},{\bf v},t)
+\frac{1}{m}{\bf F}_t({\bf x})P({\bf x},{\bf v},t)\label{currc}
\end{eqnarray}
The effective currents in position space  $\overline {\bf J_x} ({\bf x})$
and velocity space $\overline {\bf J_v} ({\bf v})$, can be found  by integrating  Eq. (\ref{fpc}) over ${\bf v}$ and ${\bf x}$  respectively to obtain the Fokker-Plank equation for the marginal probability distribution in position, ${\bf x}$ and velocity ${\bf v}$ respectively. From this, and using the divergence theorem,  the associated currents can be read off as 
\begin{eqnarray}
\overline{ \bf J_x} ({\bf x},t) &=& \int d {\bf v }\  {\bf v } P({\bf x},{\bf v},t) \label{currxint}\\
\overline{ \bf J_v} ({\bf v},t) &=& -\int d {\bf x } \left[ \frac{T\gamma}{m^2}\nabla_{\bf v} P({\bf x},{\bf v},t) + \frac{\gamma}{m}{\bf v} P({\bf x},{\bf v},t)
-\frac{1}{m}{\bf F}_t({\bf x})P({\bf x},{\bf v},t)\right].
\end{eqnarray}
In the non-equilibrium steady state, we find that the currents are given, to first order in the perturbation  by the nonconservative force as,
\begin{eqnarray}
\overline{ \bf J_x} ({\bf x}) &=& \int d {\bf v }\  {\bf v } \Delta P_0({\bf x},{\bf v})\\
\overline{ \bf J_v} ({\bf v}) &=& -\int d {\bf x } \left[ \frac{T\gamma}{m^2}\nabla_{\bf v} \Delta P_0({\bf x},{\bf v}) + \frac{\gamma}{m}{\bf v} \Delta P_0({\bf x},{\bf v})
-\frac{1}{m}{\bf F}({\bf x})P_0({\bf x},{\bf v}) + \frac{1}{m}\nabla V({\bf x})\Delta P_0({\bf x},{\bf v})\right].
\end{eqnarray}
These expressions can then be rewritten in terms of the function $K({\bf x},{\bf v})$ to give
\begin{eqnarray}
\overline{ \bf J_x} ({\bf x}) &=& \int d {\bf v }\  {\bf v } P_0({\bf x},{\bf v})K({\bf x},{\bf v})\label{defJx}\\
\overline{ \bf J_v} ({\bf v}) &=& -\int d {\bf x } \left[ \frac{T\gamma}{m^2} P_0({\bf x},{\bf v})\nabla_{\bf v} K({\bf x},{\bf v})-\frac{1}{m}{\bf F}({\bf x})P_0({\bf x},{\bf v}) + \frac{1}{m}\nabla V({\bf x}) P_0({\bf x},{\bf v})K({\bf x},{\bf v})\right].\label{defJv}
\end{eqnarray}

\begin{figure}
\begin{center}
  \includegraphics[scale=0.3]{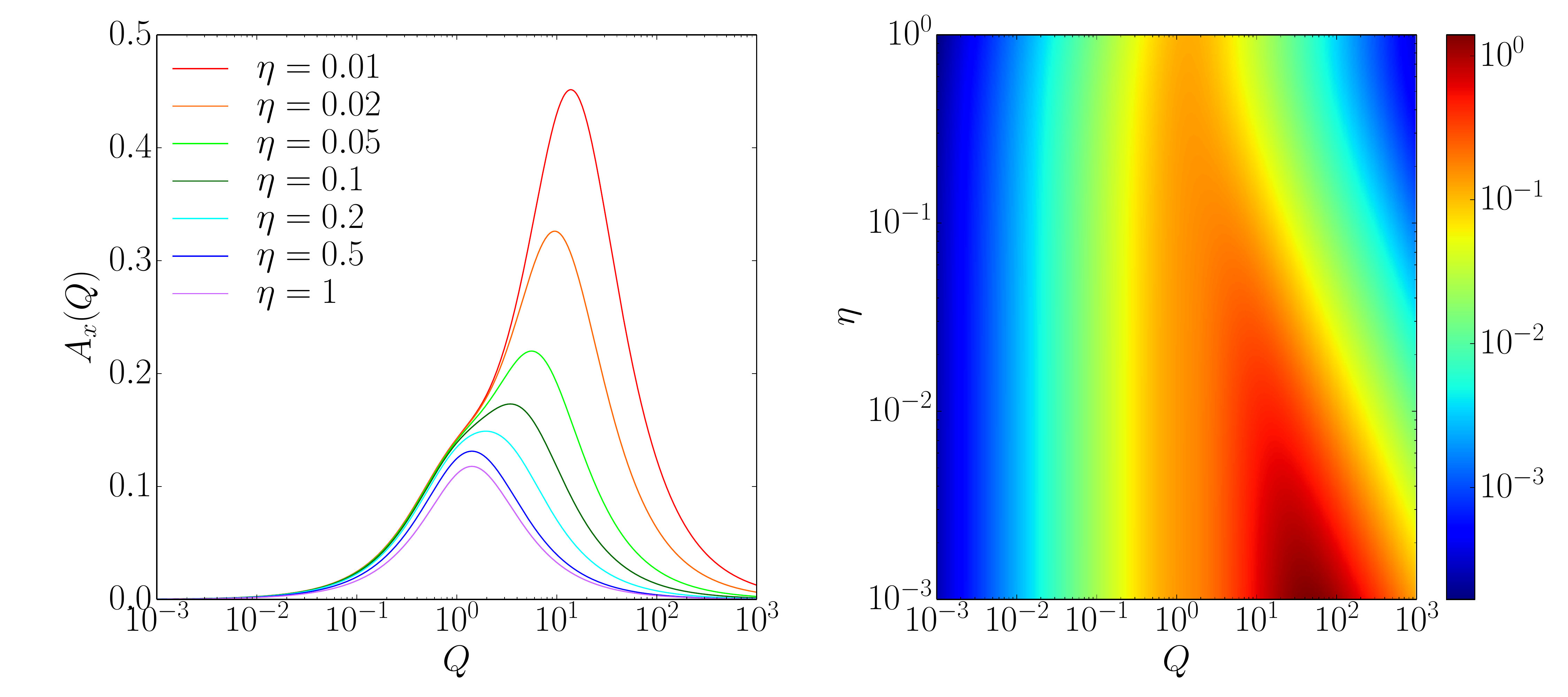}  \caption{(a) Behavior of the spatial current amplitude $A_x(\eta,Q)$, as given by Eq. (\ref{Ax})
  for various values of $\eta$. This amplitude has a single maximum for a given value $Q$, the non-equilibrium spatial current can therefore be maximized by tuning $Q$ (b) Two dimensional intensity plot of $A(\eta,Q)$. }\label{figax}
\end{center}
\end{figure}

Using the form of $K({\bf x},{\bf v})$ derived for the optical trap model we obtain 
\begin{equation}
\overline{\bf J_x} ({\bf x}) =  P_0({\bf x})\frac{4 \varepsilon \kappa \gamma}{a} \frac{ 6\gamma^2+(\eta+2)\kappa m}{[\eta\kappa m+2\gamma^2][(\eta-4)^2\kappa m + 6 (2+\eta)\gamma^2]} \left[ \eta z \rho {\bf e_\rho}+\left(\frac{2}{\beta\kappa}- \rho^2\right) {\bf e_z} \right],\label{Jx}
\end{equation}
where ${\bf e_\rho}$ is the radial unit vector in polar coordinates (recall here $\rho = \sqrt{x^2+y^2}$). In the limit $m \rightarrow 0$, we recover the result of \cite{grier2015} for
the overdamped case
\begin{equation}
\overline{\bf J_x} ({\bf x}) =  P_0({\bf x})\frac{2 \varepsilon \beta\kappa D}{a(2+\eta)} \left[  \eta z \rho {\bf e_\rho}+\left(\frac{2}{\beta\kappa}- \rho^2\right) {\bf e_z} \right],
\end{equation}
where $D$ is the effective diffusion constant, as defined by the Einstein relation, $D=T/\gamma$.
In the underdamped limit  $\gamma \rightarrow 0$, we obtain
\begin{equation}
\overline{\bf J_x} ({\bf x}) =  P_0({\bf x}) \frac{4 \varepsilon  \gamma}{a m} \frac{\eta+2}{\eta(\eta-4)^2} \left[ \eta z \rho {\bf e_\rho}+\left(\frac{2}{\beta\kappa}- \rho^2\right) {\bf e_z} \right] . 
\end{equation}
Remarkably the geometry of the current field $\overline{\bf J_x} ({\bf x})$ in space is always the same, the damping parameters only appear via an overall space independent amplitude term $A_{x}(\eta,Q)$ such that
\begin{equation}
\overline{\bf J_x} ({\bf x}) =  P_0({\bf x}) \frac{4 \varepsilon \omega_0}{a} A_x(\eta,Q) \left[ \eta z \rho {\bf e_\rho}+(\rho_c^2-\rho^2) {\bf e_z} \right],  \label{ExpJX}
\end{equation}
where $\rho_c=\sqrt{ 2/\beta \kappa}$ and 
\begin{equation}
A_x(\eta,Q) = \frac{ Q [(\eta+2)Q^2+6]}{[\eta Q^2+2][(\eta-4)^2Q^2 + 6 (2+\eta)]}. \label{Ax}
\end{equation}
We see above that when $Q\to 0$ and $Q\to\infty$, $A_x\to 0$ and thus the amplitude of the non-equilibrium current can be maximized at a given value of $Q$ when all other physical parameters are fixed,
as can be clearly seen in Fig. \ref{figax}a. We also see that the amplitude increases as $\eta$ decreases, this is to be expected as the trapping in the $z$ direction becomes weaker on reducing $\eta$.

As pointed out in \cite{grier2015}, at the point $\rho_c$ the direction of the current in the $z$ direction reverses, the value of $\rho_c$ is the value where the 
particles potential energy due to the harmonic trap attains the equilibrium value as predicted by the equipartition of energy.
We also see that 
the current in the $\rho$ direction is positive for $z>0$ and negative for $z<0$. The current in the radial (polar coordinate) direction $\rho$ is in the direction  ${\bf e_\rho}$ and its sign is proportional to $z$, thus the current flows away from the origin in the $(x,y)$ plane when $z>0$ and  towards the origin when $z<0$.  

The current in velocity space is given by $\bf v$
\begin{equation}
\overline{\bf J_v} ({\bf v}) =  
-P_{\rm MB}({\bf v}) \frac{2\varepsilon \kappa m}{a} \frac{\eta(4-\eta)\kappa m+2(\eta+2)\gamma^2}{[\eta\kappa m+2\gamma^2][(\eta-4)^2\kappa m + 6 (2+\eta)\gamma^2]} \left[ v_z (v_x {\bf e_x} + v_y {\bf e_y})  + \left(\frac{2}{\beta m} -v_\rho^2 \right) {\bf e_z} \right],
\end{equation}
where $v_\rho = \sqrt{v_x^2+v_y^2} $. In the overdamped limit, $m\rightarrow 0$, one finds
\begin{equation}
\overline{\bf J_v} ({\bf v}) = - P_{\rm MB}({\bf v}) \frac{\varepsilon \kappa m}{3a\gamma^2} \left[ v_z (v_x {\bf e_x} + v_y {\bf e_y})  + \left(\frac{2}{\beta m} -v_\rho^2 \right) {\bf e_z} \right],
\end{equation}
and the limit $\gamma\rightarrow 0$, we find 
\begin{equation}
\overline{\bf J_v} ({\bf v}) = - P_{\rm MB}({\bf v}) \frac{2\varepsilon}{a(4-\eta)} \left[ v_z (v_x {\bf e_x} + v_y {\bf e_y})  + \left(\frac{2}{\beta m} -v_\rho^2 \right) {\bf e_z} \right].
\end{equation}

\begin{figure}
\begin{center}
  \includegraphics[scale=0.3]{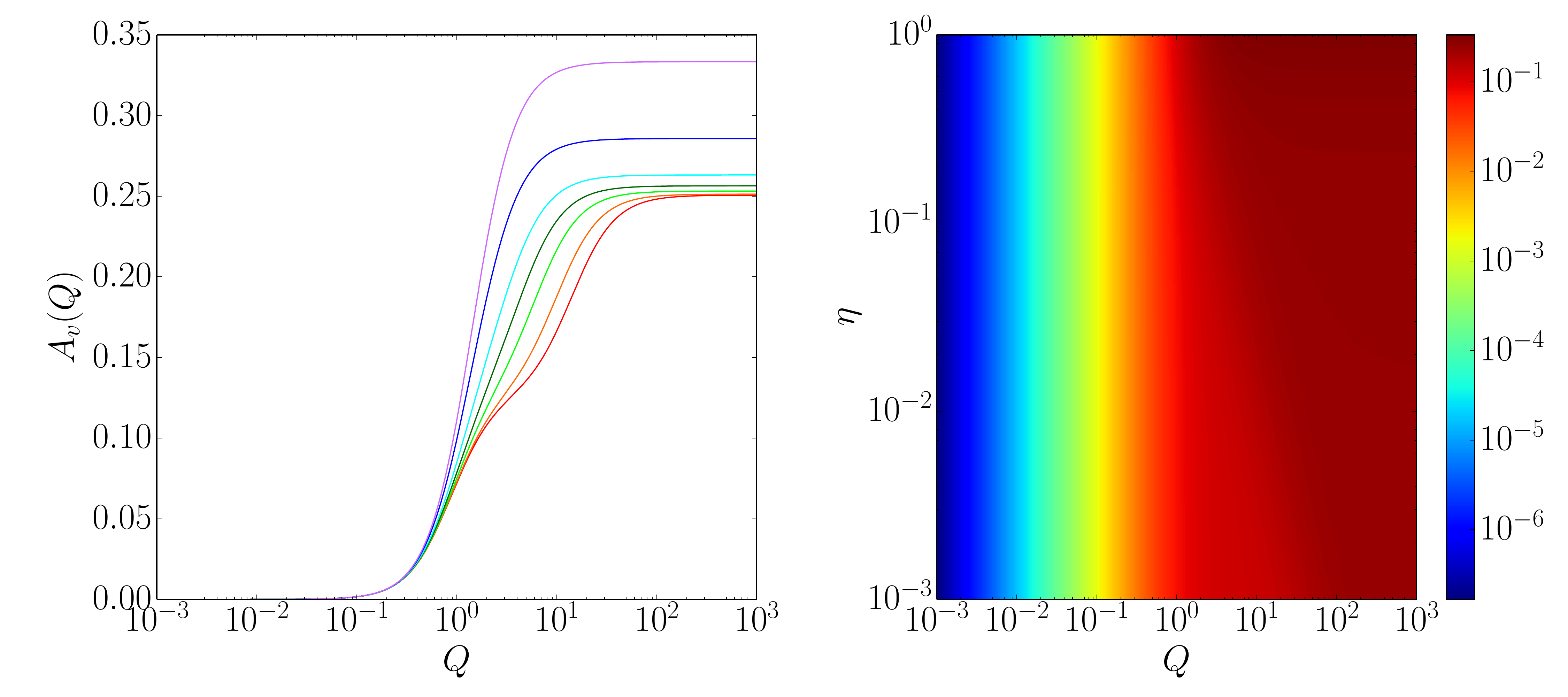}
  \caption{(a) Behavior of $A_v(\eta,Q)$, the amplitude of the current in velocity space, for several values of $\eta$. (b) Intensity plot of $A_v(\eta,Q)$.}\label{figav}
\end{center}
\end{figure}

Again we see that the geometric structure in phase space is independent of the damping 
parameters and that the damping parameters only appear in a velocity independent constant amplitude. Remarkably the  geometric structure is basically identical to that of the position current if we reverse the direction and make the correspondence ${\bf v}\equiv {\bf x}$. Here the $Q$-dependent amplitude of the current $A_v(\eta,Q)$ is defined such that

\begin{equation}
\overline{\bf J_v} ({\bf v}) =  
-P_{\rm MB}({\bf v}) \frac{2\varepsilon}{a}A_v(\eta,Q) \left[ v_z (v_x {\bf e_x} + v_y {\bf e_y})  + \left(v_c^2 -v_\rho^2 \right) {\bf e_z} \right],
\end{equation}
where $v_c=\sqrt{ 2/\beta m}$, and 
\begin{equation}
A_v(\eta,Q)= \frac{Q^2[\eta(4-\eta)Q^2+2(\eta+2)]}{[\eta Q^2+2][(\eta-4)^2Q^2 + 6 (2+\eta)]}.\label{av}
\end{equation}
Here we find that as $Q\to 0$  $A_v(\eta,Q) \to 0$, whereas $A_{v}(\eta,Q)= 1/(4-\eta)$ as  $Q\to\infty$. This function is monotonically increasing with  $Q$, as can be seen in Fig. \ref{figav}a, it reaches a plateau at large $Q$. In contrast to the amplitude of the spatial current $A_x(\eta,Q)$ we see that the amplitude is largest for large values of $\eta$, {\em i.e.} for strong trapping in the $z$ direction.

Note that the apparent divergence for $\eta=4$ is removed due to the  presence of terms of next order  in  $\gamma$, however the point $\eta =4$ should present some form of resonance in the current. The value $\eta=4$ is however far from its typical experimental value which is around $0.3$ \cite{letter}.

\subsection{Application to general particles in optical trap model with the ${\bf F}_{scat,2}$ force}

We want to find $\Delta P_s({\bf x},{\bf v}) = P_0({\bf x},{\bf v}) K'({\bf x},{\bf v})$ satisfying the equation (\ref{egamma}) with ${\bf F'}$ defined in Eq.~(\ref{fprime}) instead of ${\bf F}$. The solution of this equation must have the form
\begin{eqnarray}
\label{gammaprime}
K'({\bf x},{\bf v}) &=& z [A'+B'(x^2+y^2)+C'(xv_x+yv_y)+D'(v_x^2+v_y^2)] \nonumber\\
&+& v_z [E'+F'(x^2+y^2)+G'(xv_x+yv_y)+H'(v_x^2+v_y^2)] \nonumber\\
&+& J'z^3 + K'z^2v_z + L'zv_z^2 + M'v_z^3.
\end{eqnarray}
After simplification, the coefficients are given by
\begin{eqnarray}
A'&=&- \frac{4\varepsilon'\gamma^2}{S} [6\gamma^2+(8-5\eta)\kappa m], \\
B'&=&\frac{\beta\varepsilon'\kappa}{S} [\eta(4-\eta)(\kappa m)^2+4(4-\eta)\gamma^2\kappa m+12\gamma^4], \\
C'&=&\frac{2\eta\beta\varepsilon'\gamma\kappa m }{S}[6\gamma^2+(\eta+2)\kappa m], \\
D'&=&-\frac{\eta\beta\varepsilon'\kappa m^2}{S} [6\gamma^2+(4-\eta)\kappa m], \\
E'&=&\frac{4\varepsilon'\gamma m}{S} [6\gamma^2+(4-\eta)\kappa m], \\
F'&=&-\frac{2\beta\varepsilon'\gamma\kappa m}{S}[6\gamma^2+(\eta+2)\kappa m] \\
G'&=&- \frac{\beta\varepsilon' \kappa m^2}{S}[6(\eta-2)\gamma^2+\eta(\eta-4)\kappa m], \\
H'&=&\frac{4\beta\varepsilon \gamma\kappa m^3}{S} (\eta-1), \\
J'&=&- \frac{\beta\epsilon'\kappa}{3a}, \\
K'&=&L'=M'=0
\end{eqnarray}
where $S=a[\eta\kappa m+2\gamma^2][(\eta-4)^2\kappa m + 6 (2+\eta)\gamma^2]$. Denoting  the marginal steady state probability density function for the position by $\Delta P_s({\bf x}) = \int d{\bf v} K'({\bf x},{\bf v}) P_0({\bf x},{\bf v})$, we find from Eq.~(\ref{P0underdamped}) and (\ref{gammaprime})
\begin{equation}
\Delta P_s({\bf x}) = P_0({\bf x}) \frac{\beta \kappa \varepsilon' z}{a} \left[ \left(x^2+y^2-\frac{2}{\beta\kappa}\right) \frac{ \eta(4-\eta)(\kappa m)^2+4(4-\eta)\kappa m \gamma^2 +12\gamma^4}{[\eta\kappa m+2\gamma^2][(\eta-4)^2\kappa m + 6 (2+\eta)\gamma^2]} - \frac{z^2}{3}\right].
\end{equation}
We recover the Eq.~(\ref{pertoverprime}) taking the limit $m\rightarrow 0$. Denoting the marginal steady state probability density function for the velocity by $\Delta P_{\rm MB}'({\bf v}) = \int d{\bf x} K'({\bf x},{\bf v}) P_0({\bf x},{\bf v})$, we find from Eq.~(\ref{P0underdamped}) and (\ref{gammaprime})
\begin{equation}
\Delta P_{\rm MB}'({\bf v}) = P_{\rm MB}({\bf v}) \frac{4\beta \kappa \varepsilon' v_z}{a} \left(v_x^2+v_y^2-\frac{2}{\beta m} \right) \frac{\gamma\kappa m^3 (\eta-1)}{[\eta\kappa m+2\gamma^2][(\eta-4)^2\kappa m + 6 (2+\eta)\gamma^2]}.
\end{equation}
The effective current in position space is defined as $\overline {\bf J_x'} ({\bf x})= \int d {\bf v }\  {\bf v } P_0({\bf x},{\bf v})K'({\bf x},{\bf v})$ from Eq.~(\ref{defJx}). Using the expression of $K'$ from Eq.~(\ref{gammaprime}), we get
\begin{equation}
\overline{\bf J_x'} ({\bf x}) =  P_0({\bf x}) \frac{2 \kappa\gamma \varepsilon'}{a} \frac{ 6\gamma^2+(\eta+2)\kappa m}{[\eta\kappa m+2\gamma^2][(\eta-4)^2\kappa m + 6 (2+\eta)\gamma^2]} \left[ \eta z (x {\bf e_x}+y {\bf e_y})-\left(x^2+y^2 - \frac{2}{\beta\kappa}\right) {\bf e_z} \right].\label{Jxprime}
\end{equation}
The effective current in velocity space is defined from Eq.~(\ref{defJv}) as
\begin{equation}
\overline{ \bf J_v'} ({\bf v}) = -\int d {\bf x } \left[ \frac{T\gamma}{m^2} P_0({\bf x},{\bf v})\nabla_{\bf v} K'({\bf x},{\bf v})-\frac{1}{m}{\bf F'}({\bf x})P_0({\bf x},{\bf v}) + \frac{1}{m}\nabla V({\bf x}) P_0({\bf x},{\bf v}) K'({\bf x},{\bf v}) \right].
\end{equation}
After some simplifications, we find
\begin{equation}
\overline{\bf J_v'} ({\bf v}) =  
-P_{\rm MB}({\bf v}) \frac{\varepsilon' \kappa m}{a} \frac{\eta(4-\eta)\kappa m+2(\eta+2)\gamma^2}{[\eta\kappa m+2\gamma^2][(\eta-4)^2\kappa m + 6 (2+\eta)\gamma^2]} \left[ v_z (v_x {\bf e_x} + v_y {\bf e_y})  - \left(v_x^2+v_y^2-\frac{2}{\beta m}\right) {\bf e_z} \right].
\end{equation}
We see that the above results currents induced by the force ${\bf F}'$ are remarkably similar to those induced by the non- conservative part of the force ${\bf F}_n$ of the force ${\bf F}'$. This can in fact be seen much more easily by noting that the nonconservative parts of ${\bf F}$ and ${\bf F}'$ are proportional. Indeed, we can write 
\begin{eqnarray}
\frac{{\bf F}'}{\varepsilon'} &=& \frac{\kappa}{a}\left( \nabla [ \frac{x^2+ y^2}{2}-\frac{z^3}{3} ]- {\bf e}_z \frac{x^2 + y^2}{2}\right)\nonumber\\
&=& -\nabla U + \frac{{\bf F}}{2\varepsilon},
\end{eqnarray}
where $U = -\kappa(\frac{x^2+ y^2}{2}-\frac{z^3}{3} )$ is a potential difference between the two nonconservative forces. This potential term does not contribute to the current at the order if perturbation theory used here, thus explaining the simple relationship between the two currents found here.

\subsection{Comparison with numerical simulation}

We will now verify the analytical results for the nonequilibrium steady state probability distribution function and the associated currents, presented above, by comparing them
with the results of numerical integration of the corresponding Langevin equations which are given by
\begin{eqnarray}
m\ddot x+\gamma\dot x + \kappa_x x &=& \sqrt{2\gamma T}\xi_x \label{nlx}\\
m\ddot y+\gamma\dot y + \kappa_y y &=& \sqrt{2\gamma T}\xi_y\label{nly}\\
m\ddot z+\gamma\dot z + \kappa_z z &=& \sqrt{2\gamma T}\xi_z + \varepsilon\kappa_x a\left(1-\frac{x^2+y^2}{a^2}\right).\label{nlz}
\end{eqnarray}
Here, $\xi_x$, $\xi_y$ and $\xi_z$ are independent Gaussian white noises.
We consider the system with the following adimensionalized parameters, where $a=1$ and $\kappa_x/m = \omega_0^2 = 1$ such that
\begin{eqnarray}
\ddot x+ Q^{-1} \dot x + x &=& \sqrt{2 Q^{-1}} \sigma \xi_x \label{simux}\\
\ddot y+ Q^{-1} \dot y + \eta_y y &=& \sqrt{2 Q^{-1}} \sigma \xi_y \label{simuy}\\
\ddot z+ Q^{-1} \dot z + \eta_z z &=& \sqrt{2 Q^{-1}} \sigma \xi_z + \varepsilon \left(1-x^2-y^2\right) \label{simuz},
\end{eqnarray}
where $Q=\omega_0/\Gamma$ is the quality factor, $\sigma^2 = (\beta \kappa_x a^2)^{-1}$ and  $\eta_y = \kappa_y/\kappa_x$ and $\eta_z = \kappa_z/\kappa_x$. The numerical integration  of these equations is carried out using the algorithm of Sivak {\sl et al.} \cite{sivak2014}.
The probability density function is estimated using a spatial binning procedure, for boxes of volume $v= \Delta x \Delta y \Delta z$, as
\begin{equation}
P_0({\bf x}) = \frac{1}{v{\cal N}}\sum_{i=1}^{\cal N} I({\bf X}_{t_i}, {\bf x}),\label{pav}
\end{equation}
where $I({\bf X}_{t_i}, {\bf x})=1$ if ${\bf X}_{t_i}$, the particle position at time $t_i= i \Delta t$, is in the bin of ${\bf x}$ and ${\cal N}$ is the total number of points in the trajectory.
The spatial current is computed from Eq.~(\ref{currxint}) using the estimator 
\begin{equation}
\overline{{\bf J_{\bf x}}}({\bf x}) = \frac{1}{v\cal N}\sum_{i=1}^{\cal N} I({\bf X}_{t_i}\label{cav}
, {\bf x}) {\bf V}_{t_i},
\end{equation}
where ${\bf V}_{t_i}$ is the measured particle velocity at time $t_i$.

\begin{figure}
\begin{center}
  \includegraphics[width=18cm]{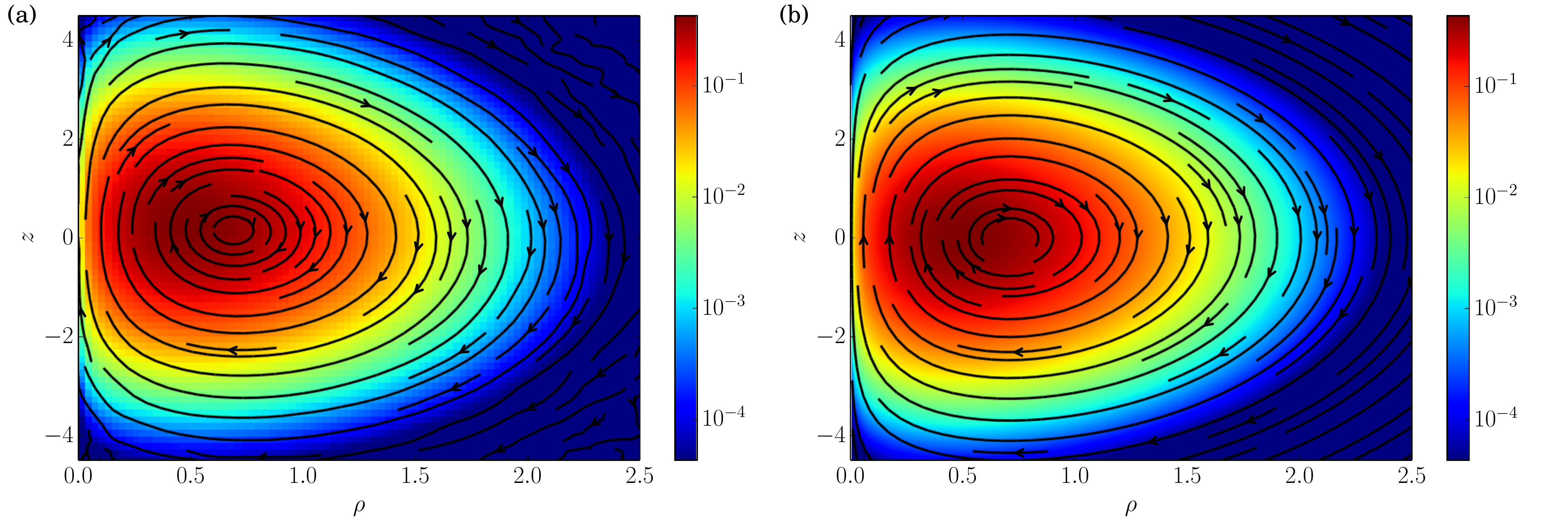}
  \caption{(a) Numerical simulation results for the stationary probability density $P_0$ obtained from the equations (\ref{simux})-(\ref{simuz}) with $\varepsilon=0.1$, $\sigma=0.5$, $\eta_y=1$, $\eta_z=0.2$ and $Q=1$ obtained from the estimator given by Eq. (\ref{pav}). The vector field  correspond to the local stochastic average of the local velocity ${\bf V}$ to obtain the stationary current $\overline{\bf J_x} (\rho , z)$ via Eq. (\ref{cav}). (b) For the same parameters, the leading order expressions of the probability density and of the current are plotted here from Eqs. (\ref{P0underdamped}) and (\ref{ExpJX}) respectively.}\label{figprob}
\end{center}
\end{figure}

The numerically obtained value of the steady state probability distribution for the parameters $\varepsilon=0.1$, $\sigma=0.5$, $\eta_y=1$, $\eta_z=0.2$ and $Q=1$ while the bins are taken with respect to $\Delta x = \Delta y = 6\sigma/100$ and $\Delta z = 6 \sigma_z/100$ (where $\sigma_z = \sigma/\sqrt{\eta}$). The simulation is carried out using for the  integration time step $\Delta t = 10^{-2}$ and over a total measurement  time $t_m = 10$ for a total of ${\cal N} = 5. 10^9$ different trajectories. We show  in  Fig \ref{figprob}(a) the estimated steady state distribution  as a grey-scale (color online)  plot, along with the vector field plot of the current (shown as arrows). In Fig. \ref{figprob}b we see the corresponding quantities predicted by the perturbation theory employed in this paper. The agreement is excellent, as it should be for the small value of $\varepsilon$ used here. 

\section{Power spectra for underdamped particles in perturbed Harmonic potentials}
 The computation of the spectral density function for the particle position in the trap model considered here has been carried out in \cite{laporta2011}. Here we generalize this calculation to the case of underdamped Brownian particles and show that significant differences arise. We then verify the calculation by comparing it with numerical simulations.
 
 \subsection{Analytical calculation}
Here we compute the power function of an underlying harmonic process subjected to  a perturbing non-conservative force. The equations of motion for the components of the particle position are again given by Eqs. (\ref{nlx}, \ref{nly}, \ref{nlz}), and in  this particular approximation only the 
steady state distribution  of $z$ is changed (the positions $x$ and $y$ are not coupled to the non-conservative force and the process $z$ is a slave to the positions $(x,y)$ \cite{slave}). The addition of the perturbation gives $z$ a non-zero average value given by
\begin{equation}
\langle z\rangle =  \frac{\varepsilon a}{\eta} \left\langle 1-\frac{x^2+y^2}{a^2}\right\rangle = \frac{\varepsilon a}{\eta}\left(1 -\frac{2T}{\kappa a^2}\right). 
\end{equation}
Due to the linearity  of Eq. (\ref{nlz}) we can decompose $z$ as
\begin{equation}
z(t) = \langle z\rangle + \zeta_{ef}(t) + \zeta_n(t),
\end{equation}
where $z_{ef}$ is the equilibrium fluctuation due to the thermal noise obeying
\begin{equation}
m\ddot \zeta_{ef}+\gamma\dot \zeta_{ef} +  \kappa \eta \zeta_{ef} = \sqrt{2\gamma T}\xi_z,
\end{equation}
and $\zeta_{n}$ is the fluctuation in $z$ due to the fluctuations of the non-conservative force, $\zeta_{n}$ then obeys
\begin{equation}
m\ddot \zeta_{n}+\gamma\dot \zeta_{n} + \kappa \eta \zeta_{n} = f_n,
\end{equation}
where 
\begin{equation}
f_n = -\frac{\varepsilon\kappa}{a}\left(x^2-\langle x^2\rangle + y^2-\langle y^2\rangle\right).\label{fn}
\end{equation}
is the fluctuating part of the non-conservative noise in the $z$ direction.
Fourier transforming in time then gives
\begin{equation}
\tilde \zeta_{ef}(\omega) = \frac{\sqrt{2T\gamma}\tilde \xi_z(\omega)}{-m\omega^2 + i\omega\gamma + \kappa\eta},
\end{equation}
and 
\begin{equation}
\tilde \zeta_{n}(\omega) = \frac{ f_n(\omega)}{-m\omega^2 + i\omega\gamma + \kappa\eta}.
\end{equation}
The power spectrum of the fluctuations $\zeta = \zeta_{ef} + \zeta_{n}$ is then given by
\begin{equation}
\langle \tilde\zeta(\omega)\tilde\zeta(\omega')\rangle = \langle \tilde\zeta_{ef}(\omega)\tilde\zeta_{ef}(\omega')\rangle 
+ \langle \tilde\zeta_{n}(\omega)\tilde\zeta_{n}(\omega')\rangle,
\end{equation}
where
\begin{equation}
\langle \tilde\zeta_{ef}(\omega)\tilde\zeta_{ef}(\omega')\rangle =2\pi \delta(\omega +\omega') S_{zz-eq}(\omega),
\end{equation}
is just given by the ordinary equilibrium fluctuation spectrum
\begin{equation}
S_{zz-eq}(\omega) = \frac{2T\gamma}{(\kappa\eta - m\omega^2)^2 + \omega^2\gamma^2}.
\end{equation}
The contribution from the non-conservative force to the power spectrum is 
\begin{equation}
\langle \tilde\zeta_{n}(\omega)\tilde\zeta_{n}(\omega')\rangle = 2\pi \delta(\omega +\omega') S_{nn}(\omega).
\end{equation}
Defining 
\begin{equation}
\langle \tilde f_n(\omega)\tilde f_n(\omega')\rangle = 2\pi\delta(\omega+\omega') C(\omega),
\end{equation}
then gives
\begin{equation}
S_{nn}(\omega) = \frac{C(\omega)}{(\kappa\eta - m\omega^2)^2 + \omega^2\gamma^2}.\label{snc}
\end{equation}
To proceed with the computation of $C(\omega)$ we use 
\begin{equation}
\widetilde{x^2}\ (\omega)= \frac{1}{2\pi}\int d\nu\ \tilde x(\omega-\nu)\tilde x(\nu),
\end{equation}
and so the associated connected correlation function is given by
\begin{equation}
\langle \widetilde{x^2}\ (\omega)\widetilde{x^2} (\omega')\rangle_c = \frac{1}{(2\pi)^2}\int d\nu d\nu' \left\{ \langle \tilde x(\omega-\nu)\tilde x(\omega'-\nu')\rangle \langle \tilde x(\nu)\tilde x(\nu')\rangle + \langle \tilde x(\omega-\nu)\tilde x(\nu')\rangle \langle \tilde x(\omega'-\nu')\tilde x(\nu)\rangle \right\}.
\end{equation}
In terms of the equilibrium spectral function for $x$ 
\begin{equation}
 S_{xx-eq}(\omega) =  \frac{2T\gamma}{(\kappa - m\omega^2)^2 + \omega^2\gamma^2}, 
\end{equation}
we thus find 
\begin{equation}
\langle \widetilde{x^2}\ (\omega)\widetilde{x^2} (\omega')\rangle_c =2\pi \delta(\omega+\omega') \int \frac{d\nu}{\pi}S_{xx-eq}(\omega-\nu) S_{xx-eq}(\nu) .
\end{equation}
Using this we obtain 
\begin{eqnarray}
C(\omega) &=& \frac{\varepsilon^2\kappa^2}{a^2}\int \frac{d\nu}{\pi}S_{xx-eq}(\omega-\nu) S_{xx-eq}(\nu) + S_{yy-eq}(\omega-\nu) S_{yy-eq}(\nu)\nonumber \\
&=& 2\frac{\varepsilon^2\kappa^2}{a^2}\int \frac{d\nu}{\pi}S_{xx-eq}(\omega-\nu) S_{xx-eq}(\nu)\label{intC},
\end{eqnarray}
the last line holding when the system is isotropic in the $(x,y)$ plane.

In order to compute $C(\omega)$ we note that the poles of $S_{xx-eq}(\nu)$ are given by
\begin{equation}
\nu_{p} = \frac{\pm i\gamma\pm \sqrt{4\kappa m -\gamma^2}}{2m}.
\end{equation}
In the underdamped case where $4\kappa m >\gamma^2$ the poles can be written as
\begin{equation}
\nu = \nu_+, \ \nu = \overline\nu_+, \ \nu = \nu_-, \ \nu = \overline\nu_-,
\end{equation}
where 
\begin{equation}
\nu_\pm = \frac{ i\gamma\pm \sqrt{4\kappa m -\gamma^2}}{2m}.
\end{equation}
In the evaluation of the integral in Eq. (\ref{intC}) the poles that contribute from $S_{xx-eq}(\nu)$ come from the upper complex plane and thus only $\nu =\nu_\pm$ contribute. The poles 
of $S_{xx-eq}(\omega -\nu)$ can be written to be  at $\nu -\omega = \nu_p$ and thus $\nu = \nu_p + \omega$. Therefore, only the poles $\nu = \nu_\pm + \omega$ contribute from 
$S(\omega -\nu)$. 
In terms of these poles we can write
\begin{eqnarray}
&&S_{xx-eq}(\nu)S_{xx-eq}(\omega -\nu) = \frac{4T^2\gamma^2}{m^4}\times \nonumber \\&&
\frac{1}{(\nu - \nu_+)(\nu - \nu_-)(\nu - \nu_+-\omega)(\nu - \nu_--\omega)(\nu - \overline\nu_+)(\nu -\overline \nu_-)(\nu - \overline \nu_+-\omega)(\nu - \overline\nu_--\omega)}.
\end{eqnarray}
Evaluating the residues in the upper complex plane we thus find
\begin{eqnarray}
&&C(\omega) = \frac{16i\varepsilon^2\kappa^2 T^2 \gamma^2}{m^4 a^2}\times \nonumber \\
&&\frac{1}{(\nu_+ - \nu_-)(-\omega)(\nu_+ - \nu_--\omega)(\nu_+ - \overline\nu_+)(\nu_+ -\overline \nu_-)(\nu_+ - \overline \nu_+-\omega)(\nu_+ - \overline\nu_--\omega)}\nonumber \\
&+&\frac{1}{(\nu_ - -\nu_+)(\nu_- - \nu_+-\omega)(-\omega)(\nu_- - \overline\nu_+)(\nu_- -\overline \nu_-)(\nu_- - \overline \nu_+-\omega)(\nu_- - \overline\nu_--\omega)}\nonumber \\
&+&\frac{1}{(\omega)(\nu_++\omega - \nu_-)(\nu_+ - \nu_-)(\nu_++\omega -\overline \nu_+)(\nu_+ +\omega - \overline \nu_-)(\nu_+ - \overline\nu_+)(\nu_+ - \overline\nu_-)}\nonumber \\
&+&\frac{1}{(\nu_-+\omega - \nu_+)(\omega)(\nu_ -- \nu_+)(\nu_-+\omega - \overline\nu_+)(\nu_-+\omega -\overline \nu_-)(\nu_ - -\overline \nu_+)(\nu_- - \overline\nu_-)}.
\end{eqnarray}
We now proceed by writing
\begin{equation}
\nu_\pm = \frac{i\Gamma \pm \Omega}{2},
\end{equation}
where $\Gamma = \gamma/m$ and $\Omega = \sqrt{4\kappa m-\gamma^2}/m$. Using this yields
\begin{eqnarray}
&&C(\omega) = \frac{16i\varepsilon^2\kappa^2 T^2 \gamma^2}{m^4 a^2}\times \nonumber \\
&&\frac{1}{\Omega(-\omega)(\Omega-\omega)i\Gamma(i\Gamma +\Omega)(i\Gamma-\omega)(i\Gamma + \Omega-\omega)}\nonumber \\
&+&\frac{1}{-\Omega(-\Omega-\omega)(-\omega)(i\Gamma-\Omega)i\Gamma(i\Gamma - \Omega-\omega)(i\Gamma-\omega)}\nonumber \\
&+&\frac{1}{\omega(\Omega+\omega )\Omega(i\Gamma+\omega )(i\Gamma +\Omega+\omega )i\Gamma(i\Gamma + \Omega)}\nonumber \\
&+&\frac{1}{(\omega - \Omega)\omega(-\Omega)(i\Gamma -\Omega+\omega )(i\Gamma+\omega )(i\Gamma-\Omega)i\Gamma}.
\end{eqnarray}
This finally gives
\begin{equation}
C(\omega) = \frac{64\varepsilon^2\kappa^2 T^2 \gamma^2}{m^4 a^2\Gamma (\Gamma^2 +\Omega^2)}
\frac{5\Gamma^2 + \omega^2 + \Omega^2}{(\Gamma^2 + \omega^2)(\Gamma^2 + (\omega-\Omega)^2)(\Gamma^2 + (\omega+\Omega)^2)} \label{comega}
\end{equation}
Substituting this into Eq. (\ref{snc}) now gives
\begin{equation}
S_{zz}(\omega)= 2T \gamma\frac{\left[ 1+\frac{32\varepsilon^2\kappa^2 T \gamma}{m^4 a^2\Gamma (\Gamma^2 +\Omega^2)}\frac{5\Gamma^2 + \omega^2 + \Omega^2}{(\Gamma^2 + \omega^2)(\Gamma^2 + (\omega-\Omega)^2)(\Gamma^2 + (\omega+\Omega)^2)}\right]}{(\kappa\eta -m\omega^2)^2 + \omega^2 \gamma^2},
\end{equation} 
which can be written as
\begin{equation}
S_{zz}(\omega)= 2T \gamma\frac{\left[ 1+\frac{8\varepsilon^2 \kappa T }{m^2  a^2}\frac{5\Gamma^2 + \omega^2 + \Omega^2}{(\Gamma^2 + \omega^2)(\Gamma^2 + (\omega-\Omega)^2)(\Gamma^2 + (\omega+\Omega)^2)}\right]}{(\kappa\eta -m\omega^2)^2 + \omega^2 \gamma^2}.
\end{equation} 
In terms of, nearly, all the physical parameters we have
\begin{equation}
S_{zz}(\omega)= \frac{2T \gamma}{(\kappa\eta -m\omega^2)^2 + \omega^2 \gamma^2}
\left[ 1 + \frac{8\varepsilon^2 T \kappa}{  a^2}\frac{4\gamma^2 + 4 \kappa m + m^2\omega^2}{(\gamma^2 + m^2\omega^2)(4\kappa-2 \omega m\Omega +m\omega^2)(4\kappa+2 \omega m\Omega +m\omega^2)}\right],
\end{equation}
where 
\begin{equation}
m\Omega = \sqrt{4\kappa m -\gamma^2}.
\end{equation}
This can also be written as
\begin{equation}
S_{zz}(\omega)= \frac{2T \gamma}{(\kappa\eta -m\omega^2)^2 + \omega^2 \gamma^2}
\left[ 1 + \frac{8\varepsilon^2 T\kappa }{  a^2}\frac{4\gamma^2 + 4 \kappa m + m^2\omega^2}{(\gamma^2 + m^2\omega^2)((4\kappa- m\omega^2)^2+4 \omega^2\gamma^2)  }\right].
\end{equation}
In the overdamped limit, where we set $m=0$, one finds
\begin{equation}
S_{zz}(\omega)= \frac{2T \gamma}{\kappa^2\eta^2  + \omega^2 \gamma^2}
\left[ 1 + \frac{8\varepsilon^2 T \kappa}{ a^2}\frac{1 }{4\kappa^2 + \omega^2\gamma^2  }\right].
\end{equation}
which agrees with the calculation for an overdamped Brownian motion in \cite{laporta2011}.

In the case where the trap is also anisotropic in the $(x,y)$ plane we write the stiffnesses in the directions $x$, $y$ and $z$ as $\kappa_x$, $\kappa_y$ and $\kappa_z$, and use the corresponding values of $w_x$ and $w_y$ for the fluctuating nonconservative force which we write from Eq. (\ref{fscat1}) as
\begin{equation}
{\bf F} = {\bf F}_x + {\bf F}_y  + \kappa \varepsilon a {\bf e}_z 
 \end{equation}
 where 
 \begin{equation} 
 \varepsilon = \frac{\kappa_z z_0(z_0k-1)}{\kappa a} 
  \end{equation}
 with $a = \sqrt{a_x a_y}$ where $a_x = w_x/\sqrt{2}$ and $a_y = w_y/\sqrt{2}$. Here $\kappa_g =\sqrt{\kappa_x\kappa_y}$ is the geometric mean of the stiffness in the directions $x$ and $y$. We can thus write
 \begin{eqnarray}
 {\bf F}_x &=& -\frac{\kappa_g\varepsilon_x}{a_x} x^2 {\bf e}_z\\
 {\bf F}_y &=& -\frac{\kappa_g\varepsilon_y}{a_y} y^2 {\bf e}_z,
 \end{eqnarray}
where $\varepsilon_x = \varepsilon a/a_x$ and $\varepsilon_y = \varepsilon a/a_y$. We can now simply use the results obtained above (as the contributions from ${\bf F}_x$ and ${\bf F}_y$ are additive) to obtain 
\begin{eqnarray}
S_{zz}(\omega)&=& \frac{2T \gamma}{(\kappa_z -m\omega^2)^2 + \omega^2 \gamma^2}
\left[ 1 + \frac{4\varepsilon^2a^2 \kappa_g^2T }{ \kappa_x a_x^4}\frac{4\gamma^2 + 4 \kappa_x m + m^2\omega^2}{(\gamma^2 + m^2\omega^2)((4\kappa_x- m\omega^2)^2+4 \omega^2\gamma^2)} \right. \nonumber \\
&+&\left.\frac{4\varepsilon^2a^2\kappa_g^2 T }{ \kappa_y a_y^4}\frac{4\gamma^2 + 4 \kappa_y m + m^2\omega^2}{(\gamma^2 + m^2\omega^2)((4\kappa_y- m\omega^2)^2+4 \omega^2\gamma^2)} \right].
\end{eqnarray} 
If we denote the geometric mean of the two waists in the $x$ and $y$ directions by
$w_g = \sqrt{w_x w_y}$, we find that in terms of the waste variables we have
\begin{eqnarray}
S_{zz}(\omega)&=& \frac{2T \gamma}{(\kappa_z -m\omega^2)^2 + \omega^2 \gamma^2}
\left[ 1 + \frac{8\varepsilon^2w_g^2 \kappa_g^2T }{ \kappa_x w_x^4}\frac{4\gamma^2 + 4 \kappa_x m + m^2\omega^2}{(\gamma^2 + m^2\omega^2)((4\kappa_x- m\omega^2)^2+4 \omega^2\gamma^2)} \right. \nonumber \\
&+&\left.\frac{8\varepsilon^2w_g^2\kappa_g^2 T }{ \kappa_y w_y^4}\frac{4\gamma^2 + 4 \kappa_y m + m^2\omega^2}{(\gamma^2 + m^2\omega^2)((4\kappa_y- m\omega^2)^2+4 \omega^2\gamma^2)} \right].
\label{powerspectrumxy}
\end{eqnarray} 

\begin{figure}
\begin{center}
  \includegraphics[width=18cm]{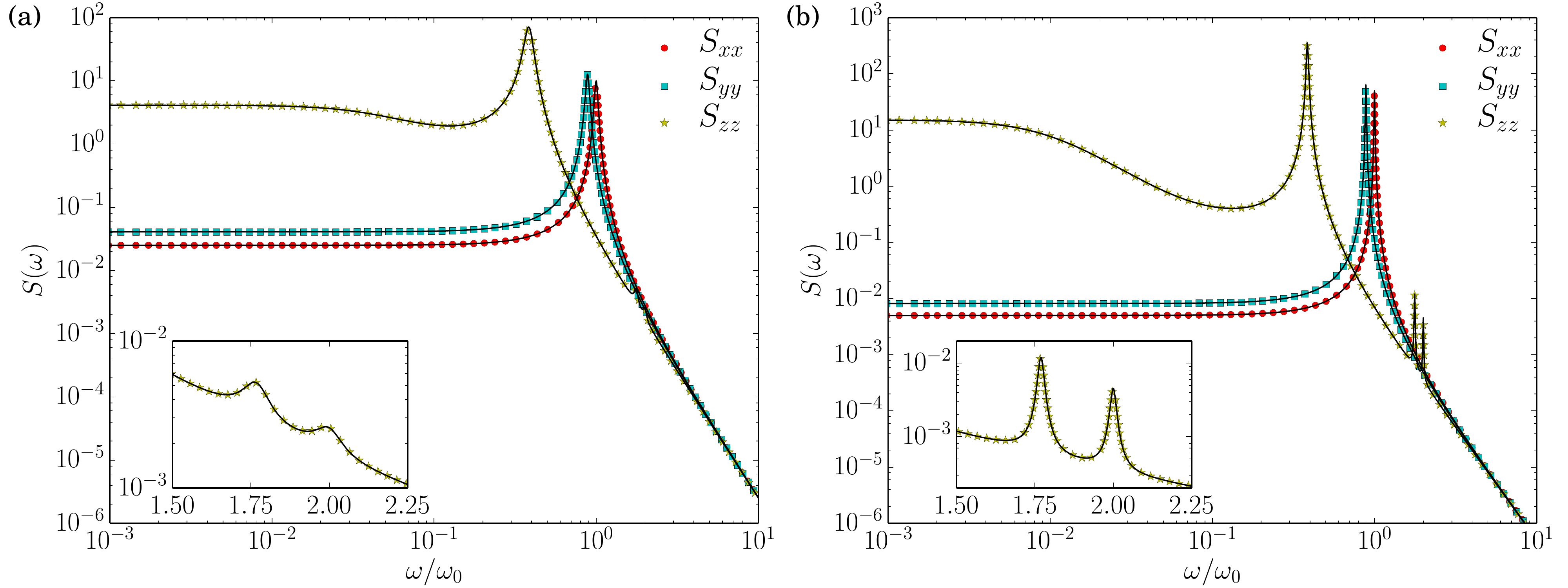}
  \caption{Power spectrum for underdamped particles in presence of scattering, obtain from numerical integration of equations (\ref{simux})-(\ref{simuz}). The numerical simulation are plotted with symbols: $S_{xx}$ with disks, $S_{yy}$ with squares and $S_{zz}$ with stars. The lines represent the exact equations for these functions, given by the general formula (\ref{powerspectrumxy}). The parameters taken here are $\eta_y=0.7828$, $\eta_z=0.1485$, $\sigma=0.5$, $\varepsilon=0.1$ and $Q=20$ for (a) and $Q=100$ for (b). In insets, a zoom over the two peaks of $S_{zz}$ at $\omega = 2 \omega_0$ and $\omega = 2 \sqrt{\eta_y} \omega_0 \simeq 1.77 \omega_0$ is represented.}\label{figscattering}
\end{center}
\end{figure}

If we consider the low frequency limit $\omega\to 0$ and consider the simplest case where the system is isotropic in the $(x,y)$ plane, {\em i.e. $w_x=w_y$}, we find
\begin{equation}
S_{zz}(0)= \frac{2T \gamma}{\kappa^2\eta^2}
\left[ 1 + \frac{2\varepsilon^2 T}{ a^2}\frac{\gamma^2 +  \kappa m }{\gamma^2\kappa }\right].
\end{equation}
which has the form
\begin{equation}
S_{zz}(0)= \frac{2T}{\kappa^2\eta^2}\left(\gamma A + \frac{B}{\gamma}\right),
\end{equation}
where $A$ and $B$ are damping independent constants given by
\begin{equation}
A= 1 + \frac{2\varepsilon^2 T }{ \kappa a^2} ,\  B = \frac{2\varepsilon^2 Tm }{ a^2}.
\end{equation}
This predicts that $S_z(0)$ attains a minimal value at $\gamma_c = \sqrt{B/A}
$. 

\subsection{Signatures of time reversal symmetry breaking}

The crucial difference between equilibrium and non-equilibrium systems stems from the breaking of time reversal symmetry in the latter systems. It is this symmetry breaking that leads to the appearance of currents in non-equilibrium steady states \cite{kubo}.  Here we show time reversal symmetry breaking can be inferred from temporal measurements of the particle positions.

Consider two observables $A$ and $B$ in a steady state, their correlation function is defined by
\begin{equation}
C_{AB}(t) = \langle A(t) B(0)\rangle = \langle A(0) B(-t)\rangle =C_{BA}(-t),
\end{equation}
where the last step above uses the fact that any steady state by definition must be invariant by time translations \cite{kubo}.

The time reversal symmetry is broken when
\begin{equation}
C_{BA}(t)\neq C_{BA}(-t)
\end{equation}
so that in this case the Onsager reciprocal relation \cite{kubo},
\begin{equation}
C_{AB}(t) = C_{BA}(t),
\end{equation}
does not hold. 

For an  equilibrium system, as $C_{AB}(t)=C_{BA}(t)$, we find that  $\tilde C_{AB}(\omega)$ is thus real. In the problem studied here we need to find two operators $A$ and $B$, with non-zero correlation to demonstrate the violation of time reversal symmetry. The obvious choice $A= z$ and 
$B=x$ (or $B=y$) is not useful as the corresponding correlators are
zero. However choosing $B= x^2$ (or $B=y^2$) does yield a non-zero correlation function. However, to simplify the 
calculations which follow, we will  choose $B= f_n(t)$, the fluctuating part of the component of the nonconservative force in the $z$ direction, which, see Eq. (\ref{fn}),  is a function of $x^2$ and $y^2$ with zero mean, and from which the aforementioned correlators can trivially extracted. We define 
the correlation function
\begin{equation}
C_{zf_n}(t) = \langle z(t) f_n(0)\rangle,
\end{equation}
which simplifies to 
\begin{equation}
C_{zf_n}(t) = \langle \zeta_n(t) f_n(0)\rangle,
\end{equation}
because $\zeta_{ef}$ is independent of $f_n$ and $f_n$ is chosen to have zero mean.  
This now gives
\begin{equation}
\tilde C_{zf_n}(\omega)= \frac{C(\omega)}{-m\omega^2 + i\gamma\omega +\kappa\eta},
\end{equation}
where $C(\omega)$ is given by Eq.(\ref{comega}). As $C(\omega)$ is
real, we see that $\tilde C_{zf_n}(\omega)$ is not real and has an imaginary component
\begin{equation}
\tilde C''_{zf_n}(\omega)= \frac{-\gamma\omega C(\omega)}{(\kappa\eta -m\omega^2)^2 + \gamma^2\omega^2 } \label{corrzf}
\end{equation}

Therefore if one can experimentally measure the imaginary component of this correlation function, one has another method of 
probing the non-equilibrium steady state of trapped particles in optical 
traps.

\subsection{Numerical simulations}

Following the equations (\ref{simux})-(\ref{simuz}), where we adimensionalized the space and time variables by writing them in units of $a$ and $\omega_0 = \sqrt{\kappa_x/m}$ respectively. We compute  the correlation function $\langle X(\omega) Y(\omega') \rangle = 2 \pi \delta(\omega + \omega') S_{XY} (\omega)$ from a Fast Fourier Transform of the time variable functions $X(t)$ and $Y(t)$ over $2^{27} \sim 10^8$ points and $10^3$ stochastic averages where after discretization $S_{XY} (\omega_i) = \langle X(\omega_i) Y^*(\omega_i) \rangle$ (where $.^*$ represents the complex conjugate). We verify here the exact equation for $S_{zz}$ as predicted in Eq. (\ref{powerspectrumxy}) and the time reversal breaking predicted in  via the  non-zero value of $C''_{zf}$ function as is given in Eq. (\ref{corrzf}).

\begin{figure}
\begin{center}
  \includegraphics[width=18cm]{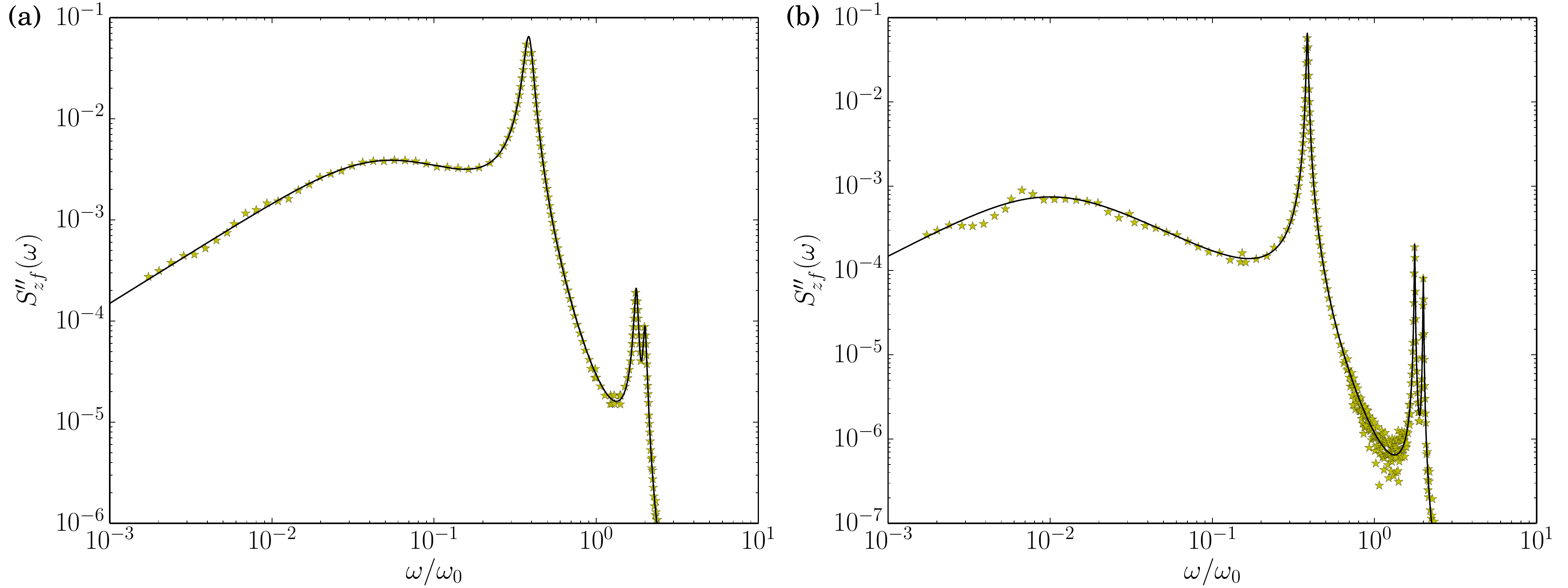}
  \caption{The imaginary part of correlation function $S_{zf}(\omega)$ for the parameters $\eta_y=0.7828$, $\eta_z=0.1485$, $\sigma=0.5$, $\varepsilon=0.1$ and $Q=20$ (a) and $Q=100$ (b). The lines represent the exact equation (\ref{corrzf}) and stars represent the numerical simulations with a stochastic average over 1000 realizations.}\label{figtimereversal}
\end{center}
\end{figure}

\section{Conclusion}
Optical trapping can be carried out in both liquid and gas phases. In
the gas phase, controlling the pressure can be used to modify the friction coefficient of the trapped particle as the gas viscosity changes.
The fact that the system is subject to a non-conservative force means that the resulting steady state depends on the precise form of the dynamics, in this case the damping coefficient $\gamma$. We have shown that this dependence on $\gamma$ can be demonstrated by measuring currents in the steady state, notably the currents associated with the marginal probability densities in both position and velocity space. Futhermore, the signature of optical scattering can be seen in the behavior of the spectral density function of the position and is particularly strong at low frequencies. We have also derived a correlation function between the particle position and the non-conservative force (equivalent to examining correlations between the position $Z(t)$ and the variable $X^2(t)$) which has a component which does not respect time reversal symmetry, another measurement of non-equilibrium behavior.

Many of the non-equilibrium results derived here have a non-monotonic dependence on the particle's friction coefficient and are thus susceptible to being optimized to maximize their potential experimental signals, pointing to directions for future experimental studies.

A promising future direction of this study is to study the full nonlinear model for the optical trap to see what modifications are engendered by non-linear effects. However our numerical studies of the non-linear model (see accompanying paper \cite{letter}) suggest that the effects predicted by the harmonic model here persist, and are thus robust. Finally, even within the harmonic model here, the non-linear form of the conservative force means that we have only computed the first two moments of the temporal correlation functions and our results for the modified steady state distribution are perturbative. However it would of be interest to compute the full distribution function even for this simple harmonic model.

\begin{acknowledgments}
This work was partially funded by the
Bordeaux IdEX program - LAPHIA (ANR-10-IDEX-03-02) and the ANR project RaMaTRaF.
\end{acknowledgments}

 \end{document}